%% file: main.tex
\documentclass[11pt,onecolumn,a4paper,journal]{IEEEtran}
\usepackage{amsmath,amsfonts,amssymb}
\usepackage[a4paper,margin=1in]{geometry}
\usepackage[utf8]{inputenc}
\usepackage[T1]{fontenc}
\usepackage{graphicx}
\usepackage[dvipsnames]{xcolor}
\colorlet{RED}{red}
\colorlet{BLUE}{blue}
\colorlet{GREEN}{green}
\usepackage{array}
\usepackage{textcomp}
\usepackage{url}
\usepackage{verbatim}
\usepackage{booktabs}
\usepackage{algorithm}
\usepackage{algpseudocode}
\usepackage{subcaption}
\usepackage{orcidlink}
\usepackage[per-mode=symbol]{siunitx}
\DeclareSIUnit{\dBm}{dBm}
\DeclareSIUnit{\dB}{dB}
\usepackage[nomain, toc, numberedsection, acronym]{glossaries}
\loadglsentries{abbr.tex}
\makeglossaries
\glsdisablehyper 
\usepackage[backend=biber,style=ieee]{biblatex}
\AtBeginBibliography{\footnotesize}
\usepackage{hyperref}
\hypersetup{
  colorlinks=true,
  citecolor=blue,
  linkcolor=black,
  urlcolor=black,
  filecolor=black,
  pdfborder={0 0 0}
}
\usepackage[capitalise]{cleveref}

\newcommand{\vect}[1]{\boldsymbol{\mathrm{#1}}}
\newcommand{\mat}[1]{\boldsymbol{\mathrm{#1}}}
\newcommand{\norm}[1]{\left\lVert#1\right\rVert}

\begin{document}

\title{When Distortion Helps: Secure GNN Precoding with Nonlinear Power Amplifiers}


\author{Reza Ghasemi Alavicheh~\orcidlink{0009-0002-3620-5481},
Thomas Feys~\orcidlink{0000-0002-5369-4670},
Md Arifur Rahman~\orcidlink{0000-0003-3861-9290},
Fran\c{c}ois Rottenberg~\orcidlink{0000-0002-2150-8511}
\thanks{R. Ghasemi Alavicheh is with the Research and Innovation Department, IS-Wireless, Piaseczno, Poland (e-mail: r.ghasemi@is-wireless.com), and also with ESAT-DRAMCO, KU Leuven, Ghent, Belgium (e-mail: reza.ghasemialavicheh@kuleuven.be).}
\thanks{T. Feys and F. Rottenberg are with ESAT-DRAMCO, KU Leuven, Ghent, Belgium (e-mail: \{thomas.feys, francois.rottenberg\}@kuleuven.be).}
\thanks{M. A. Rahman is with Research and Innovation Department, IS-Wireless, Piaseczno, Poland (e-mail: a.rahman@is-wireless.com).}
\thanks{This work has received funding from the European Union's Horizon 2022 research and innovation program under Grant Agreement No 101120332 (EMPOWER-6G project).}}

\maketitle

\begin{abstract}
\Gls{pls} provides information-theoretic protection against eavesdropping. While existing techniques assume ideal linear transmitters, \glspl{pa} in practice introduce nonlinear distortion, typically considered detrimental to signal quality. This paper demonstrates that such distortion can instead be exploited as a security asset by redirecting it toward eavesdroppers, particularly in the power-efficient \gls{pa} saturation regime. To this end, we propose a \gls{gnn}-based precoding framework for multi-user \gls{miso} wiretap channels that maximizes the sum secrecy rate by exploiting \gls{pa} nonlinearity. {{Since the resulting optimization is highly non-convex, classical methods are intractable. The GNN instead learns precoding strategies directly from legitimate users' channel data, requiring neither eavesdropper CSI nor dedicated \gls{an} power allocation. For this, the Bussgang decomposition and a high-order polynomial \gls{pa} model provide an analytical secrecy rate as the training objective.}} At 22~dB \gls{snr} under severe \gls{pa} saturation with \gls{ibo}~$= -1$~dB, the proposed \gls{gnn} achieves 39.89\% and 35.26\% higher sum secrecy rate over \gls{mrt} and \gls{zf}, respectively, 17.99\% over \gls{an}-aided \gls{mrt} and 8.67\% over \gls{an}-aided \gls{zf}, with 58.13--75.31\% lower standard deviation across all baselines.
\end{abstract}
\input{sections/01_introduction}
\input{sections/02_system_model}
\input{sections/03_sndr_formulation}

\input{sections/04_proposed_gnn}

\input{sections/05_baseline_methods}
\input{sections/06_numerical_results}
\input{sections/07_complexity}
\input{sections/08_conclusion}

\printglossaries
\printbibliography
\appendices
\input{sections/09_appendix}

\end{document}

%% file: abbr.tex

\newacronym{2d}{2D}{two-dimensional}
\newacronym{3d}{3D}{three-dimensional}
\newacronym{3gpp}{3GPP}{3rd generation partnership project}
\newacronym{3msk}{3MSK}{3-level minimum-shift-keying}
\newacronym{4g}{4G}{fourth-generation}
\newacronym{5g}{5G}{fifth-generation}
\newacronym{6dof}{6DoF}{six degrees of freedom}
\newacronym{6g}{6G}{sixth-generation}
\newacronym{6tisch}{6TiSCH}{IPv6 over the TSCH mode of IEEE 802.15.4e}
\newacronym{abd}{ABD}{ambient-backscattering device}
\newacronym{abp}{ABP}{authentication by personalisation}
\newacronym{ac}{AC}{alternating current}
\newacronym{aci}{ACI}{adjacent channel interference}
\newacronym{aclr}{ACLR}{adjacent channel leakage ratio}
\newacronym{acpr}{ACPR}{adjacent channel power ratio}
\newacronym{adc}{ADC}{analog-to-digital converter}
\newacronym{adr}{ADR}{adaptive data rate}
\newacronym{aec}{AEC}{aluminum electrolytic capacitor}
\newacronym{aes}{AES}{Advanced Encryption Standard}
\newacronym{afa}{AFA}{adaptive frequency agility}
\newacronym{afe}{AFE}{analog front end}
\newacronym{ag}{AG}{array gain}
\newacronym{agc}{AGC}{automatic gain controller}
\newacronym{agps}{A-GPS}{Assisted Global Positioning System} 
\newacronym{ai}{AI}{artificial intelligence}
\newacronym{aimd}{AIMD}{Additive-Increase/Multiplicative-Decrease}
\newacronym{aiot}{A-IoT}{Ambient IoT}
\newacronym{alk}{ALK}{Alkaline}
\newacronym{am}{AM}{amplitude modulation}
\newacronym{amam}{AM/AM}{amplitude modulation to amplitude modulation}
\newacronym{ambc}{AmBC}{ambient backscatter communication}
\newacronym{amc}{AMC}{automatic modulation classification}
\newacronym{amp}{AMP}{approximate message passing}
\newacronym{ampm}{AM/PM}{amplitude modulation to phase modulation}
\newacronym{an}{AN}{artificial noise}
\newacronym{aoa}{AOA}{angle-of-arrival}
\newacronym{aod}{AOD}{angle-of-departure}
\newacronym{ap}{AP}{access point}
\newacronym{api}{API}{application program interface}
\newacronym{apt}{APT}{acoustic power transfer}
\newacronym{apu}{APU}{access point unit}
\newacronym{ar}{AR}{augmented reality}
\newacronym{arima}{ARIMA}{Auto-Regressive Integrated Moving Average}
\newacronym{arp}{ARP}{Antenna Reference Point}
\newacronym{asic}{ASIC}{application specific integrated circuit}
\newacronym{ask}{ASK}{amplitude-shift keying}
\newacronym{at}{AT}{ATtention}
\newacronym{auv}{AUV}{autonomous underwater vehicle}
\newacronym{awg}{AWG}{arbitrary waveform generator}
\newacronym{awgn}{AWGN}{additive white Gaussian noise}
\newacronym{baw}{BAW}{bulk acoustic wave}
\newacronym{bb}{BB}{base-band}
\newacronym{bc}{BC}{backscatter communication}
\newacronym{bcjr}{BCJR}{Bahl-Cocke-Jelinek-Raviv}
\newacronym{bd}{BD}{backscatter device}
\newacronym{be}{BE}{Belgium}
\newacronym{ber}{BER}{bit error rate}
\newacronym{bf}{BF}{beamforming}
\newacronym{bga}{BGA}{ball grid array}
\newacronym{bibc}{BiBC}{bistatic backscatter communication}
\newacronym{bjt}{BJT}{bipolar junction transistor}
\newacronym{bldc}{BLDC}{brushless DC}
\newacronym{ble}{BLE}{Bluetooth Low Energy}
\newacronym{bler}{BLER}{block error rate}
\newacronym{bod}{BOD}{Brown-Out Detection}
\newacronym{bom}{BOM}{bill of materials}
\newacronym{bpsk}{BPSK}{binary phase-shift keying}
\newacronym{brp}{BRP}{beam refinement process}
\newacronym{bs}{BS}{base station}
\newacronym{btwt}{bTWT}{broadcast TWT}
\newacronym{bu}{BU}{booster unit}
\newacronym{bw}{BW}{bandwidth}
\newacronym{ca}{CA}{carrier emitter}
\newacronym{cad}{CAD}{channel activity detection}
\newacronym{cars}{CARS}{calibration reference signal}
\newacronym{cbm}{CBM}{condition based maintenance}
\newacronym{cbra}{CBRA}{contention-based random access}
\newacronym{cc}{CC}{constant current}
\newacronym{ccdf}{CCDF}{complementary cumulative distribution function}
\newacronym{ccnn}{CCNN}{circular convolutional neural network}
\newacronym{ccs}{CCS}{correlative channel sounder}
\newacronym{cdf}{CDF}{cumulative distribution function}
\newacronym{cdma}{CDMA}{code division-multiple access}
\newacronym{cdrx}{CDRX}{connected mode DRX}
\newacronym{ce}{CE}{coverage enhancement}
\newacronym{cea}{CEA}{Controlled Environment Agriculture}
\newacronym{ced}{CED}{cumulative energy density}
\newacronym{ceofdm}{CE-OFDM}{constant-envelope OFDM}
\newacronym{cept}{CEPT}{European Conference of Postal and Telecommunications Administrations}
\newacronym{cf}{CF}{cell-free}
\newacronym{cfmmimo}{CF-mMIMO}{cell-free massive MIMO}
\newacronym{cfo}{CFO}{carrier frequency offset}
\newacronym{cfra}{CFRA}{contention-free random access}
\newacronym{cir}{CIR}{channel impulse response}
\newacronym{cla}{CLA}{closed-loop approach}
\newacronym{clk}{CLK}{clock}
\newacronym{cmos}{CMOS}{complementary metal oxide semiconductor}
\newacronym{cn}{CN}{Core network}
\newacronym{cnn}{CNN}{convolutional neural network}
\newacronym{co}{CO}{combinatorial optimization}
\newacronym{cordic}{CORDIC}{coordinate rotation digital computer}
\newacronym{cost}{COST}{commercial off-the-shelf}
\newacronym{cots}{COTS}{commercial off-the-shelf}
\newacronym{cp}{CP}{cyclic prefix}
\newacronym{cpe}{CPE}{common phase error}
\newacronym{cpfsk}{CPFSK}{continuous phase frequency shift keying}
\newacronym{cpm}{CPM}{continuous phase modulation}
\newacronym{cpo}{CPO}{carrier phase offset}
\newacronym{cpt}{CPT}{capacitive power transfer}
\newacronym{cpu}{CPU}{central-processing unit}
\newacronym{cpw}{CPW}{coplanar waveguide}
\newacronym{cqi}{CQI}{channel quality indicator}
\newacronym{cr}{CR}{coding rate}
\newacronym{crc}{CRC}{cyclic redundancy check}
\newacronym{crlb}{CRLB}{Cram\'er-Rao lower bound}
\newacronym{crs}{CRS}{cell reference signal}
\newacronym{crtwt}{C-RTWT}{coordinated restricted TWT}
\newacronym{cs}{CS}{compressed sensing}
\newacronym{cse}{CSE}{channel state estimation}
\newacronym{csi}{CSI}{channel state information}
\newacronym{csp}{CSP}{contact service point}
\newacronym{css}{CSS}{chirp spread spectrum}
\newacronym{cu}{CU}{central unit}
\newacronym{cv}{CV}{constant voltage}
\newacronym{cw}{CW}{continuous wave}
\newacronym{d2d}{D2D}{device-to-device}
\newacronym{d2r}{D2R}{device-to-reader}
\newacronym{dab}{DAB}{Digital Audio Broadcasting}
\newacronym{dac}{DAC}{digital-to-analog converter}
\newacronym{daq}{DAQ}{data acquisition system}
\newacronym{das}{DAS}{distributed antenna systems}
\newacronym{db}{DB}{duty-cycled barebone}
\newacronym{dbpsk}{DBPSK}{Differential Binary Phase Shift Keying}
\newacronym{dc}{DC}{direct current}
\newacronym{dcc}{DCC}{dynamic cooperation clustering}
\newacronym{ddc}{DDC}{digital down conversion}
\newacronym{de}{DE}{drain efficiency}
\newacronym{de-ewma}{DE-EWMA}{Dynamic Error-Exponentially Weighted Moving Average}
\newacronym{dect}{DECT}{Digital Cordless Telecommunications}
\newacronym{dft}{DFT}{discrete Fourier transform}
\newacronym{dftsofdm}{DFT-s-OFDM}{discrete Fourier Transform spread OFDM}
\newacronym{dftsofdmfdss}{DFT-s-OFDM-FDSS}{DFT-s-OFDM with frequency-domain spectral shaping}
\newacronym{dftsofdmfdssse}{DFT-s-OFDM-FDSS-SE}{DFT-s-OFDM with FDSS and spectral extension}
\newacronym{dl}{DL}{downlink}
\newacronym{dlc}{DLC}{Distributed Laser Charging}
\newacronym{dli}{DLI}{direct link interference}
\newacronym{dlpf}{DLPF}{Dual Differential Low Pass Filter}
\newacronym{dlt}{DLT}{Distributed Ledger Technology}
\newacronym{dm}{DM}{diffuse multipath}
\newacronym{dma}{DMA}{Direct Memory Access}
\newacronym{dmac}{DMAC}{Direct Memory Access Controller}
\newacronym{dmc}{DMC}{diffuse multipath component}
\newacronym{dmimo}{D-MIMO}{distributed MIMO}
\newacronym{dnn}{DNN}{deep neural network}
\newacronym{doa}{DOA}{direction-of-arrival}
\newacronym{dp}{DP}{differencial privacy}
\newacronym{dpd}{DPD}{digital pre-distortion}
\newacronym{dpdk}{DPDK}{Data Plane Development Kit}
\newacronym{dram}{DRAM}{dynamic random-access memory}
\newacronym{drcs}{$\Delta$RCS}{differential-radar cross section}
\newacronym{drx}{DRX}{Discontinuous Reception Mode}
\newacronym{dsb}{DSB}{double-sideband}
\newacronym{dsl}{DSL}{digital subscriber line}
\newacronym{dsp}{DSP}{digital signal processing}
\newacronym{dss}{DSS}{Dataset Storage Standard}
\newacronym{dsss}{DSSS}{direct-sequence spread spectrum}
\newacronym{duc}{DUC}{digital up-converter}
\newacronym{dvb}{DVB}{Digital Video Broadcasting}
\newacronym{e2e}{E2E}{end-to-end}
\newacronym{easa}{EASA}{European Union Aviation Safety Agency}
\newacronym{ebg}{EBG}{electromagnetic bandgap}
\newacronym{ebw}{EBW}{excess bandwidth}
\newacronym{ec}{EC}{European Commission}
\newacronym{ecc}{ECC}{elliptic curve cryptography}
\newacronym{ecdf}{eCDF}{empirical cumulative distribution function}
\newacronym{ecdlp}{ECDLP}{elliptic curve discrete logarithm problem}
\newacronym{ecsp}{ECSP}{edge computing service point}
\newacronym{edlc}{EDLC}{electrostatic double-layer capacitors}
\newacronym{edrx}{eDRX}{Extended Discontinuous Reception Mode}
\newacronym{ee}{EE}{energy efficiency}
\newacronym{egprs}{EGPRS}{Enhanced Data Rates for GSM Evolution}
\newacronym{eh}{EH}{energy harvesting}
\newacronym{eipm}{EIPM}{Energy Intelligence Platform Module}
\newacronym{eirp}{EIRP}{equivalent isotropically radiated power}
\newacronym{em}{EM}{electromagnetic}
\newacronym{embb}{eMBB}{enhanced Mobile Broadband}
\newacronym{en}{EN}{energy neutral}
\newacronym{end}{END}{energy-neutral device}
\newacronym{enob}{ENOB}{effective number of bits}
\newacronym{eol}{EoL}{end of life}
\newacronym{ep}{EP}{energy profiler}
\newacronym{epd}{EPD}{electronic paper display}
\newacronym{epu}{EPU}{edge processing unit}
\newacronym{er}{ER}{Energy Receiver}
\newacronym{erc}{ERC}{European Radiocommunications Committee}
\newacronym{erp}{ERP}{effective radiation power}
\newacronym[plural=ESCs,firstplural=electronic speed controllers (ESCs)]{esc}{ESC}{electronic speed control}
\newacronym{esd}{ESD}{electrostatic discharge}
\newacronym{esl}{ESL}{electronic shelf label}
\newacronym{esr}{ESR}{equivalent series resistance}
\newacronym{et}{ET}{Energy Transmitter}
\newacronym{etsi}{ETSI}{European Telecommunications Standards Institute}
\newacronym{evd}{EVD}{eigenvalue decomposition}
\newacronym{evm}{EVM}{Error Vector Magnitude}
\newacronym{ewlb}{eWLB}{Embedded Wafer Level Ball Grid Array}
\newacronym{ewma}{EWMA}{Exponentially Weighted Moving Average}
\newacronym{fa}{FA}{federation anchor}
\newacronym{fair}{FAIR}{Findability, Accessibility, Interoperability, and Reuse of digital assets}
\newacronym{fc}{FC}{fusion center}
\newacronym{fcc}{FCC}{Federal Communications Commission}
\newacronym{fcf}{FCF}{frequency correlation function}
\newacronym{fd}{FD}{front-haul distance}
\newacronym{fdd}{FDD}{frequency-division duplexing}
\newacronym{fde}{FDE}{frequency domain equalizer}
\newacronym{fdm}{FDM}{frequency-division multiplexing}
\newacronym{fdma}{FDMA}{frequency division multiple access}
\newacronym{fdss}{FDSS}{frequency-domain spectral shaping}
\newacronym{fec}{FEC}{forward-error correction}
\newacronym{fem}{FEM}{finite element analysis}
\newacronym{fembb}{feMBB}{further enhanced mobile broadband}
\newacronym{fft}{FFT}{fast Fourier transform}
\newacronym{fh}{FH}{fronthaul}
\newacronym{fhss}{FHSS}{frequency hopping spread spectrum}
\newacronym{fifo}{FIFO}{First In, First Out}
\newacronym{fim}{FIM}{Fisher information matrix}
\newacronym{fits}{FITS}{Flexible Image Transport System}
\newacronym{fl}{FL}{federated learning}
\newacronym{fom}{FoM}{figure of merit}
\newacronym{flop}{FLOP}{floating-point operation}
\newacronym{fov}{FOV}{field of view}
\newacronym{fpga}{FPGA}{field-programmable gate array}
\newacronym{fqt}{FQT}{Fully Quantized Training}
\newacronym{fr1}{FR1}{frequency range 1}
\newacronym{fr2}{FR2}{frequency range 2}
\newacronym{fram}{FRAM}{Ferroelectric Random Access Memory}
\newacronym{fsk}{FSK}{frequency shift keying}
\newacronym{fspl}{FSPL}{free space path loss}
\newacronym{fss}{FSS}{frequency selective surface}
\newacronym{gan}{GaN}{gallium nitride}
\newacronym{gb}{GB}{grant-based}
\newacronym{gdpr}{GDPR}{general data protection regulation}
\newacronym{gf}{GF}{grant-free}
\newacronym{gmsk}{GMSK}{Gaussian minimum-shift keying}
\newacronym{gnb}{gNB}{Next Generation Node B}
\newacronym{gni}{GNI}{gross national income}
\newacronym{gnn}{GNN}{graph neural network}
\newacronym{gnss}{GNSS}{global navigation satellite system}
\newacronym{gpclk}{GPCLK}{general purpose clock}
\newacronym{gpio}{GPIO}{General-Purpose Input/Output}
\newacronym{gpl}{GPL}{GNU General Public License}
\newacronym{gprs}{GPRS}{General Packet Radio Services}
\newacronym{gps}{GPS}{Global Positioning System}
\newacronym{gpu}{GPU}{graphical processing unit}
\newacronym{gr}{GR}{GNU~Radio}
\newacronym{grc}{GRC}{GNU Radio Companion}
\newacronym{gscm}{GSCM}{geometry‐based stochastic model}
\newacronym{gsm}{GSM}{Global System for Mobile Communications}  %
\newacronym{gspm}{GSpM}{Generalized spatial modulation}
\newacronym{gwp}{GWP}{Global Warming Potential}
\newacronym{har}{HAR}{human activity recognition}
\newacronym{harq}{HARQ}{hybrid automatic repeat request}
\newacronym{hat}{HAT}{hardware attached on top}
\newacronym{hcs}{HCS}{human-centric services}
\newacronym{hdf5}{HDF5}{Hierarchical Data Format version 5}
\newacronym{hdr}{HDR}{high data rate}
\newacronym{hf}{HF}{high frequency}
\newacronym{hfss}{HFSS}{High Frequency Simulator Software}
\newacronym{hmd}{HMD}{head-mounted display}
\newacronym{hpbm}{HPBM}{half power beam width}
\newacronym{hpbw}{HPBW}{half-power beamwidth}
\newacronym{HyMPRo}{HyMPRo}{Hybrid Multi-Path Routing algorithm}
\newacronym{i.i.d.}{i.i.d.}{independent and identically distributed}
\newacronym{i2c}{I2C}{Inter-Integrated Circuit}
\newacronym{iaq}{IAQ}{Indoor Air Quality}
\newacronym{ib}{IB}{in-band}
\newacronym{ibbc}{IBBC}{inter-band beam configuration}
\newacronym{ibo}{IBO}{input back-off}
\newacronym{ic}{IC}{integrated circuit}
\newacronym{iccs}{ICCS}{Ilmsens correlative channel sounder}
\newacronym{ici}{ICI}{intercarrier interference}
\newacronym{icnirp}{ICNIRP}{International Commission on Non-Ionizing Radiation Protection}
\newacronym{id}{ID}{information decoding}
\newacronym{idft}{IDFT}{inverse discrete Fourier transform}
\newacronym{idxm}{IDXM}{index modulation}
\newacronym{ieee}{IEEE}{Institute of Electrical and Electronics Engineers}
\newacronym{if}{IF}{intermediate-frequency}
\newacronym{ifft}{IFFT}{inverse fast-Fourier-transform}
\newacronym{iid}{i.i.d.}{independently and identically distributed}
\newacronym{iis}{IIS}{integrated information system}
\newacronym{im}{IM}{intermodulation}
\newacronym{imd}{IMD}{intermodulation distortion}
\newacronym{imu}{IMU}{inertial measurement unit}
\newacronym{inh}{InH}{indoor hotspot office}
\newacronym{io}{IO}{input/output}
\newacronym{ioe}{IoE}{Internet of Everything}
\newacronym{iot}{IoT}{Internet of Things}
\newacronym{ipt}{IPT}{inductive power transfer}
\newacronym{ipy}{IPY}{Interventions per Year}
\newacronym{iq}{IQ}{in-phase and quadrature}
\newacronym{iqi}{IQI}{IQ imbalance}
\newacronym{ir}{IR}{infrared}
\newacronym{isi}{ISI}{intersymbol interference}
\newacronym{ism}{ISM}{industrial, scientific and medical}
\newacronym{isp}{ISP}{internet service provider}
\newacronym{itu}{ITU}{International Telecommunication Union}
\newacronym{itu-r}{ITU-R}{ITU – Radiocommunication Sector}
\newacronym{jesd}{JESD}{Joint Electron Devices Engineering Council}
\newacronym{jfet}{JFET}{junction field effect transistor}
\newacronym{keh}{KEH}{kinetic energy harvester}
\newacronym{kpi}{KPI}{key performance indicator}
\newacronym{ktofdm}{KT-DFT-s-OFDM}{known-tail-DFT-s-OFDM}
\newacronym{kvi}{KVI}{key value indicator}
\newacronym{larva}{LARVA}{LARge Virtual Array}
\newacronym{lbt}{LBT}{listen-before-talk}
\newacronym{lca}{LCA}{life cycle assessment}
\newacronym{lch}{LCH}{logical channel}
\newacronym{lco}{LCO}{lithium cobalt oxide}
\newacronym{ldo}{LDO}{Low-dropout voltage regulator}
\newacronym{ldpc}{LDPC}{low-density parity-check}
\newacronym{ldr}{LDR}{low data rate}
\newacronym{led}{LED}{Light Emitting Diode}
\newacronym{less}{LESS}{Low Energy Scheduler Solution}
\newacronym{lev}{LEV}{light electric vehicle}
\newacronym{lfp}{LFP}{lithium iron phosphate}
\newacronym{lib}{LIB}{Lithium-Ion Battery}
\newacronym{lic}{LIC}{lithium-ion capacitor}
\newacronym{lid}{LID}{Lithium Iron Disulfide}
\newacronym{lidar}{LiDAR}{light detection and ranging}
\newacronym{liion}{Li-ion}{lithium-ion}
\newacronym{lipo}{LiPo}{lithium polymer}
\newacronym{lis}{LIS}{large intelligent surface}
\newacronym{llh}{LLH}{log-likelihood}
\newacronym{llr}{LLR}{log-likelihood ratio}
\newacronym{lls}{LLS}{link-level simulation}
\newacronym{lmd}{LMD}{Lithium Manganese Dioxide}
\newacronym{lmmse}{LMMSE}{least minimum mean square error}
\newacronym{lmo}{LMO}{lithium ion manganese oxide}
\newacronym{lna}{LNA}{low-noise amplifier}
\newacronym{lo}{LO}{local oscillator}
\newacronym{lora}{LoRa}{long range}
\newacronym{lorawan}{LoRaWAN}{long-range wide-area network}
\newacronym{los}{LoS}{line-of-sight}
\newacronym{lp}{LP}{linear programming}
\newacronym{lpf}{LPF}{low-pass filter}
\newacronym{lpt}{LPT}{laser power transfer}
\newacronym{lpwa}{LPWA}{Low Power Wide Area}
\newacronym{lpwan}{LPWAN}{low-power wide-area network}
\newacronym{lpwans}{LPWANs}{Low-Power Wide-Area Networks}
\newacronym{lqi}{LQI}{link quality indicator}
\newacronym{lrelu}{LReLU}{leaky rectified linear unit}
\newacronym{lrt}{LRT}{likelihood-ratio test}
\newacronym{ls}{LS}{least squares}
\newacronym{lsa}{LSA}{large synthetic array}
\newacronym{lsf}{LSF}{large-scale fading}
\newacronym{lsfc}{LSFC}{large-scale fading component}
\newacronym{lstm}{LSTM}{Long Short-Term Memory}
\newacronym{ltc}{LTC}{lithium thionyl chloride}
\newacronym{lte}{LTE}{Long Term Evolution}
\newacronym{lti}{LTI}{linear time-invariant}
\newacronym{lto}{LTO}{lithium titanate}
\newacronym{lusta}{LUSTA 5G }{Logistique mUltimodale Sécuritaire Téléopérée \& Autonome 5G}
\newacronym{m2m}{M2M}{machine to machine}
\newacronym{ma}{MA}{Movable Antennas}
\newacronym{mac}{MAC}{Medium Access Control}
\newacronym{mate}{MATE}{millimeter-wave MIMO testbed}
\newacronym{mc}{MC}{Monte Carlo}
\newacronym{mcl}{MCL}{Maximum Coupling Loss}
\newacronym{mcs}{MCS}{modulation and coding scheme}
\newacronym{mcu}{MCU}{microcontroller unit}
\newacronym{mec}{MEC}{multi-access edge computing}
\newacronym{mems}{MEMS}{micro-electromechanical systems}
\newacronym{mf}{MF}{matched filter}
\newacronym{milp}{MILP}{Mixed Integer Linear Programming}
\newacronym{mimo}{MIMO}{multiple-input multiple-output}
\newacronym{miso}{MISO}{multiple-input single-output}
\newacronym{ml}{ML}{machine learning}
\newacronym{mlp}{MLP}{multilayer perceptron}
\newacronym{mmic}{MMIC}{monolithic microwave integrated circuit}
\newacronym{mmimo}{mMIMO}{massive MIMO}
\newacronym{mmse}{MMSE}{minimum mean square error}
\newacronym{mmtc}{mMTC}{massive machine-typed communication}
\newacronym{mmwave}{mmWave}{millimeter wave}
\newacronym{mn}{MN}{matching network}
\newacronym{mobc}{MoBC}{monostatic backscatter communication}
\newacronym{mosfet}{MOSFET}{metal-oxide semiconductor field effect transistor}
\newacronym{mpc}{MPC}{multipath component}
\newacronym{mpp}{MPP}{magnetic power profile}
\newacronym{mppt}{MPPT}{maximum power point tracking}
\newacronym{mqtt}{MQTT}{Message Queuing Telemetry Transport}
\newacronym{mr}{MR}{maximum ratio}
\newacronym{mrc}{MRC}{maximum ratio combining}
\newacronym{mrc_em}{MRC}{maximum ratio combining}
\newacronym{mrc_EM}{MRC}{Magnetic Resonance Coupling}
\newacronym{mrt}{MRT}{maximum ratio transmission}
\newacronym{mse}{MSE}{mean square error}
\newacronym{msk}{MSK}{Minimum-Shift Keying}
\newacronym{mtc}{MTC}{Machine-Type Communication}
\newacronym{multi-rat}{Multi-RAT}{multiple radio access technology}
\newacronym{multirat}{Multi-RAT}{Multiple Radio Access Technology}
\newacronym{music}{MUSIC}{MUltiple SIgnal Classification}
\newacronym{nas}{NAS}{Neural Architecture Search}
\newacronym{navauwall}{NAVAUWALL}{AUtomated NAVigation in WALLonia}
\newacronym{nb}{NB}{narrowband}
\newacronym{nbiot}{NB-IoT}{narrowband IoT}
\newacronym{nca}{NCA}{nickel cobalt aluminum}
\newacronym{netcdf}{NetCDF}{Network Common Data Form}
\newacronym{nf}{NF}{noise figure}
\newacronym{nfc}{NFC}{near-field communication}
\newacronym{nfv}{NFV}{network function virtualization}
\newacronym{ngmn}{NGMN}{Next Generation Mobile Networks }
\newacronym{ni}{NI}{National Instruments}
\newacronym{nicd}{NiCd}{nikkel cadmium}
\newacronym{nimh}{NiMH}{nikkel metal hydride}
\newacronym{nlos}{NLoS}{non-line-of-sight}
\newacronym{nmc}{NMC}{nickel manganese cobalt}
\newacronym{nmos}{nMOS}{n-channel metal-oxide semiconductor}
\newacronym{nn}{NN}{neural network}
\newacronym{nnls}{NNLS}{non-negative least squares}
\newacronym{noma}{NOMA}{non-orthogonal multiple access}
\newacronym{np}{NP}{Neyman-Pearson}
\newacronym{npbch}{NPBCH}{Narrowband Physical Broadcast Channel}
\newacronym{npss}{NPSS}{Narrow Band Primay Synchronization Signal}
\newacronym{nr}{NR}{New Radio}
\newacronym{nrs}{NRS}{Narrow Band Reference Signal}
\newacronym{nsss}{NSSS}{Narrowband Secondary Synchronization Signal}
\newacronym{ntp}{NTP}{network time protocol}
\newacronym{oai}{OAI}{OpenAirInterface} 
\newacronym{obw}{OBW}{occupied bandwidth}
\newacronym{ofa}{OFA}{Once-for-All}
\newacronym{ofdm}{OFDM}{orthogonal frequency-division multiplexing}
\newacronym{ofdma}{OFDMA}{orthogonal frequency-division multiple access}
\newacronym{ofdmim}{OFDM-IM}{OFDM with index modulation}
\newacronym{olos}{OLoS}{obstructed-line-of-sight}
\newacronym{oma}{OMA}{orthogonal multiple access}
\newacronym{oob}{OOB}{out-of-band}
\newacronym{ook}{OOK}{on-off keying}
\newacronym{opbo}{OPBO}{output power backoff}
\newacronym{oran}{O-RAN}{open radio-access network}
\newacronym{os}{OS}{operating system}
\newacronym{ota}{OTA}{over-the-air}
\newacronym{otaa}{OTAA}{over-the-air authentication}
\newacronym{otaml}{OTA-TinyML}{Over-the-air TinyML}
\newacronym{ova}{OVA}{One-Versus-All}
\newacronym{p1}{P1}{Phase 1}
\newacronym{p2}{P2}{Phase 2}
\newacronym{p2p}{P2P}{point-to-point}
\newacronym{pa}{PA}{power amplifier}
\newacronym{pae}{PAE}{power-added efficiency}
\newacronym{pam}{PAM}{pulse amplitude modulation}
\newacronym{pana}{PanA}{Panel A}
\newacronym{panb}{PanB}{Panel B}
\newacronym{papr}{PAPR}{peak-to-average power ratio}
\newacronym{pb}{PB}{power beacon}
\newacronym{pc}{PC}{pilot count}
\newacronym{pcb}{PCB}{printed circuit board}
\newacronym{pcg}{PCG}{power consumption gain}
\newacronym{pcie}{PCIe}{Peripheral Component Interconnect Express}
\newacronym{pcsi}{PCSI}{perfect channel state information}
\newacronym{pd}{PD}{powered device}
\newacronym{pdcch}{PDCCH}{physical downlink control channel}
\newacronym{pdcp}{PDCP}{packet data convergence protocol}
\newacronym{pdf}{PDF}{probability density function}
\newacronym{pdp}{PDP}{power delay profile}
\newacronym{pdsch}{PDSCH}{physical downlink shared channel}
\newacronym{pdu}{PDU}{Protocol Data Unit}
\newacronym{pe}{PE}{processing element}
\newacronym{peb}{PEB}{positioning error bound}
\newacronym{peet}{PET}{Polyethylene terephthalate}
\newacronym{per}{PER}{packet error rate}
\newacronym{pet}{PET}{privacy enhancing technology}
\newacronym{pg}{PG}{path gain}
\newacronym{pgd}{PGD}{proximal gradient descent}
\newacronym{phy}{PHY}{physical}
\newacronym{pin}{PIN}{positive-intrinsic-negative}
\newacronym{pki}{PKI}{public key infrastucture}
\newacronym{pl}{PL}{path loss}
\newacronym{pla}{PLA}{physically large array}
\newacronym{pla2}{PLA}{polylactic acid}
\newacronym{pll}{PLL}{phase-locked loop}
\newacronym{pls}{PLS}{physical layer security}
\newacronym[plural=PMs,firstplural=person months (PMs)]{pm}{PM}{person month}
\newacronym{pmf}{PMF}{polymer microwave fiber}
\newacronym{pmic}{PMIC}{power management integrated circuit}
\newacronym{pmu}{PMU}{power management unit}
\newacronym{pn}{PN}{pseudo-noise}
\newacronym{po}{PO}{phase offset}
\newacronym{poc}{PoC}{proof of concept}
\newacronym{poe}{PoE}{power-over-Ethernet}
\newacronym{pps}{1PPS}{pulse per second}
\newacronym{pr}{PR}{phase reversal}
\newacronym{prb}{PRB}{Physical Resource Block}
\newacronym{prbs}{PRBs}{Physical Resource Blocks}
\newacronym{prs}{PRS}{Peripheral Reflex System}
\newacronym{ps}{PS}{Processing System}
\newacronym{psd}{PSD}{power spectral density}
\newacronym{pse}{PSE}{power sourcing equipment}
\newacronym{psk}{PSK}{phase shift keying}
\newacronym{psm}{PSM}{power saving mode}
\newacronym{pss}{PSS}{primary synchronisation signal}
\newacronym{ptp}{PTP}{precision-time protocol}
\newacronym{ptrs}{PTRS}{Phase-Tracking Reference Signals}
\newacronym{ptw}{PTW}{paging time window}
\newacronym{pv}{PV}{photovoltaic}
\newacronym{pw}{PW}{planar wavefront}
\newacronym{pwm}{PWM}{pulse width modulation}
\newacronym{qam}{QAM}{quadrature amplitude modulation}
\newacronym{qos}{QoS}{quality-of-service}
\newacronym{qpsk}{QPSK}{quadrature phase-shift keying}
\newacronym{qrrls}{QR-RLS}{QR decomposition based recursive least squares}
\newacronym{quadriga}{QuaDRiGa}{QUAsi Deterministic RadIo channel GenerAtor}
\newacronym{r2d}{R2D}{reader-to-device}
\newacronym{ra}{RA}{Random Access}
\newacronym{ram}{RAM}{random-access memory}
\newacronym{ran}{RAN}{radio access network}
\newacronym{rar}{RAR}{Random Access Response}
\newacronym{rat}{RAT}{radio access technology}
\newacronym{raw}{RAW}{restricted access window}
\newacronym{rb}{Rb}{Rubidium}
\newacronym{rbs}{RBS}{radio base station}
\newacronym{rbw}{RBW}{resolution bandwidth}
\newacronym{rc}{RC}{raised-cosine}
\newacronym{rcs}{RCS}{radar cross section}
\newacronym{rdl}{RDL}{redistribution layer}
\newacronym{re}{RE}{radio element}
\newacronym{red}{RED}{radio equipment directiv}
\newacronym{redcap}{RedCap}{reduced capability}
\newacronym{relu}{ReLU}{rectified linear unit}
\newacronym{rf}{RF}{radio frequency}
\newacronym{rfeh}{RFEH}{radio frequency energy harvesting}
\newacronym{rfic}{RFIC}{radio-frequency integrated circuit}
\newacronym{rfid}{RFID}{radio frequency identification}
\newacronym{rfpt}{RFPT}{radio frequency power transfer}
\newacronym{rfsoc}{RFSoC}{Radio Frequency System-on-Chip}
\newacronym{rfwpt}{RF-WPT}{radio frequency wireless power transfer}
\newacronym{rir}{RIR}{room impulse response}
\newacronym{ris}{RIS}{reflective intelligent surface}
\newacronym{rl}{RL}{reinforcement learning}
\newacronym{rlc}{RLC}{Radio Link Control}
\newacronym{rllmtc}{RLLMTC}{reliable low latency machine type communication}
\newacronym{rls}{RLS}{recursive least squares}
\newacronym{rms}{RMS}{root-mean-square}
\newacronym{rmse}{RMSE}{root-mean-square error}
\newacronym{rmt}{RMT}{random matrix theory}
\newacronym{rof}{RoF}{radio-over-fiber}
\newacronym{ros}{ROS}{robot operating system}
\newacronym{rr}{RR}{ridge regression}
\newacronym{rpi}{RPi}{Raspberry Pi}
\newacronym{rrc}{RRC}{Radio Resource Connection}
\newacronym{rreq}{RREQ}{route request packet}
\newacronym{rsrp}{RSRP}{Reference Signals Received Power}
\newacronym{rsrq}{RSRQ}{Reference Signal Received Quality}
\newacronym{rss}{RSS}{received signal strength}
\newacronym{rssi}{RSSI}{received signal strength indicator}
\newacronym{rtc}{RTC}{real time clock}
\newacronym{rtf}{RTF}{reader talks first}
\newacronym{rtk}{RTK}{real time kinematics}
\newacronym{rts}{RTS}{ray tracing simulator}
\newacronym{rtwt}{rTWT}{restricted TWT}
\newacronym{ru}{RU}{Radio Unit}
\newacronym{ruc}{rUC}{representative Use Cases}
\newacronym{rv}{RV}{random variable}
\newacronym{rw}{RW}{RadioWeaves}
\newacronym{rx}{RX}{receiver}
\newacronym{rzf}{RZF}{regularized zero forcing}
\newacronym{s-parameter}{S-parameter}{scattering parameter}
\newacronym{sa}{SA}{synchronization anchor}
\newacronym{sdar}{SDAR}{signal-to-distortion-plus-artificial ratio}
\newacronym{sa5}{SA}{Stand Alone}
\newacronym{sar}{SAR}{specific absorption rate}
\newacronym{sbl}{SBL}{sparse Bayesian learning}
\newacronym{sc}{SC}{single carrier}
\newacronym{scfde}{SC-FDE}{single-carrier modulation with frequency-domain-equalization}
\newacronym{scfdma}{SCFDMA}{single-carrier frequency division multiple access}
\newacronym{scomp}{SC}{Split Computing}
\newacronym{scs}{SCS}{sub-carrier spacing}
\newacronym{sdap}{SDAP}{service data adaptation protocol}
\newacronym{sdg}{SDG}{Sustainable Development Goal}
\newacronym{sdm}{SDM}{sigma-delta modulator}
\newacronym{sdma}{SDMA}{spatial-division multiple access}
\newacronym{sdn}{SDN}{software-defined network}
\newacronym{sdo}{SDO}{Standards Development Organization}
\newacronym{sdof}{SDoF}{sigma-delta over fiber}
\newacronym{sdr}{SDR}{software-defined radio}
\newacronym{se}{SE}{spectral efficiency}
\newacronym{sei}{SEI}{specific emitter identification}
\newacronym{ser}{SER}{symbol-error rate}
\newacronym{sf}{SF}{spreading factor}
\newacronym{sfn}{SFN}{single frequency network}
\newacronym{sfo}{SFO}{sampling frequency offset}
\newacronym{sfp}{SFP}{small form-factor pluggable}
\newacronym{sha}{SHA}{Secure Hash Algorithm}
\newacronym{SigMF}{SigMF}{Signal Metadata Format}
\newacronym{simo}{SIMO}{single-input multiple-output}
\newacronym{sinr}{SINR}{signal-to-interference-plus-noise ratio}
\newacronym{siso}{SISO}{single-input single-output}
\newacronym{slam}{SLAM}{simultaneous localization and mapping}
\newacronym{slc}{SLC}{spatial leakage suppression}
\newacronym{slerp}{SLERP}{spherical linear interpolation}
\newacronym{sma}{SMA}{SubMiniature version A}
\newacronym{smc}{SMC}{specular multipath component}
\newacronym{smps}{SMPS}{switched mode power supply}
\newacronym{smt}{SMT}{Surface Mount Technology}
\newacronym{sndr}{SNDR}{signal-to-noise-and-distortion ratio}
\newacronym{snidr}{SNIDR}{signal-to-noise-and-interference-and-distortion ratio}
\newacronym{snidar}{SNIDAR}{signal-to-noise-and-interference-and-distortion-and-AN-leakage ratio}
\newacronym{snir}{SNIR}{signal-to-interference-plus-noise ratio}
\newacronym{snr}{SNR}{signal-to-noise ratio}
\newacronym{soc}{SoC}{state of charge}
\newacronym{SoC}{SOC}{System on Chip}
\newacronym{sota}{SotA}{state of the art}
\newacronym{sp}{SP}{service point}
\newacronym{spdt}{SPDT}{single pole double throw}
\newacronym{spi}{SPI}{Serial Peripheral Interface}
\newacronym{spst}{SPST}{single pole single throw}
\newacronym{sram}{SRAM}{static random-access memory}
\newacronym{srd}{SRD}{short-range device}
\newacronym{srls}{SRLS}{standard recursive least squares}
\newacronym{srs}{SRS}{Sounding Reference Signal}
\newacronym{ssb}{SSB}{synchronisation signal block}
\newacronym{ssd}{SSD}{solid state drive}
\newacronym{ssq}{SSQ}{simulator sickness questionnaire}
\newacronym{sta}{STA}{station}
\newacronym{steam}{STEAM}{science, technology, engineering, the arts, and mathematics}
\newacronym{svd}{SVD}{singular value decomposition}
\newacronym{svm}{SVM}{support vector machine}
\newacronym{sw}{SW}{spherical wavefront}
\newacronym{swipt}{SWIPT}{simultaneous wireless information and power transfer}
\newacronym{synce}{SyncE}{Synchronous Ethernet}
\newacronym{t2t}{T2T}{tag-to-tag}
\newacronym{ta}{TA}{timing advance}
\newacronym{tau}{TAU}{tracking area update}
\newacronym{tcer}{TCER}{transported to consumed energy ratio}
\newacronym{tcp}{TCP}{Transmission Control Protocol}
\newacronym{tcxo}{TCXO}{temperature compensated crystal oscillator}
\newacronym{tdce}{TD-CE}{time-domain compression and expansion}
\newacronym{tdd}{TDD}{time division duplexing}
\newacronym{tdma}{TDMA}{time division-multiple access}
\newacronym{tdoa}{TDOA}{time-difference-of-arrival}
\newacronym{teg}{TEG}{Thermoelectric Generator}
\newacronym{teng}{TENG}{Triboelectric nanogenerators}
\newacronym{thz}{THz}{Terahertz}
\newacronym{tinyml}{TinyML}{Tiny Machine Learning}
\newacronym{tinyol}{TinyOL}{Tiny Online Learning}
\newacronym{tinyrl}{TinyRL}{Tiny Reinforcement Learning}
\newacronym{tinytl}{TinyTL}{Tiny Transfer Learning}
\newacronym{tl}{TL}{Transfer Learning}
\newacronym{to}{TO}{timing offset}
\newacronym{toa}{TOA}{time-of-arrival}
\newacronym{tof}{ToF}{time-of-flight}
\newacronym{tosm}{TOSM}{through-open-short-match}
\newacronym{tpms}{TPMS}{Tire-Pressure Monitoring System}
\newacronym{trl}{TRL}{technology readyness level}
\newacronym{trp}{TRP}{Transmission Reception Point}
\newacronym{tsg}{TSG}{Technical Specification Group}
\newacronym{tsn}{TSN}{time-sensitive networking}
\newacronym{ttf}{TTF}{tag talks first}
\newacronym{ttff}{TTFF}{Time To First Fix}
\newacronym{tti}{TTI}{transmission time interval}
\newacronym{ttm}{TTM}{time to market}
\newacronym{ttn}{TTN}{The Things Network}
\newacronym{twt}{TWT}{Target Wake Time}
\newacronym{tx}{TX}{transmitter}
\newacronym{uart}{UART}{Universal Asynchronous Receiver/Transmitter}
\newacronym{uav}{UAV}{unmanned aerial vehicle}
\newacronym{uc}{UC}{use case}
\newacronym{ucie}{UCIe}{Universal Chiplet Interconnect Express}
\newacronym{udp}{UDP}{User Datagram Protocol}
\newacronym{ue}{UE}{user equipment}
\newacronym{ugv}{UGV}{unmanned ground vehicle}
\newacronym{uhd}{UHD}{USRP hardware driver}
\newacronym{uhf}{UHF}{ultra-high frequency}
\newacronym{ul}{UL}{uplink}
\newacronym{ula}{ULA}{uniform linear array}
\newacronym{ule}{ULE}{ultra low energy}
\newacronym{ulp}{ULP}{ultra-low power}
\newacronym{uMIMO}{$\mu$-MIMO}{ultra-massive Multiple-Input Multiple-Output}
\newacronym{ummtc}{umMTC}{ultra massive machine type communication}
\newacronym{un}{UN}{United Nations}
\newacronym{unow}{uNOW}{unified non-orthogonal waveform}
\newacronym{up}{UP}{ultra passive}
\newacronym{upa}{UPA}{uniform planar array}
\newacronym{ura}{URA}{uniform rectangular array}
\newacronym{urllc}{URLLC}{ultra-reliable low-latency communications}
\newacronym{usrp}{USRP}{universal software radio peripheral}
\newacronym{uv}{UV}{unmanned vehicle}
\newacronym{uw}{UW}{unique word}
\newacronym{uwb}{UWB}{ultrawideband}
\newacronym{uwofdm}{UW-DFT-s-OFDM}{unique word DFT-s-OFDM}
\newacronym{v2v}{V2V}{vehicle-to-vehicle}
\newacronym{vco}{VCO}{voltage-controlled oscillator}
\newacronym{vep}{VEP}{virtual edge platform}
\newacronym{vf}{VF}{Vertical Farming}
\newacronym{vlc}{VLC}{visible light communication}
\newacronym{vlp}{VLP}{visible light positioning}
\newacronym{vna}{VNA}{vector network analyzer}
\newacronym{voc}{VOC}{Voltatile Organic Compound}
\newacronym{votable}{VOTable}{Virtual Observatory Table}
\newacronym{vr}{VR}{virtual reality}
\newacronym{vswr}{VSWR}{Voltage Standing Wave Ratio}
\newacronym{vuca}{VUCA}{volatile, uncertain, complex and ambiguous}
\newacronym{wb}{WB}{wideband}
\newacronym{wban}{WBAN}{wireless body area network}
\newacronym{wcma}{WCMA}{weather conditioned moving average}
\newacronym{wd}{WD}{Wireless distance}
\newacronym{wifi}{Wi-Fi}{Wireless Fidelity}
\newacronym{wimax}{WiMAX}{Worldwide Interoperability for Microwave Access}
\newacronym{wirelesshart}{WirelessHART}{Wireless Highway Addressable Remote Transducer Protocol}
\newacronym{wlan}{WLAN}{wireless LAN}
\newacronym[plural=WPs,firstplural=work packages (WPs)]{wp}{WP}{work package}
\newacronym{wpc}{WPC}{Wireless Power Consortium}
\newacronym{wpt}{WPT}{wireless power transfer}
\newacronym{wr}{WR}{White Rabbit}
\newacronym{wrsn}{WRSN}{wireless rechargeable sensor network}
\newacronym{wsn}{WSN}{wireless sensor network}
\newacronym{wur}{WuR}{wake-up radio}
\newacronym{wus}{WuS}{wake-up signal}
\newacronym{xets}{XETS}{cross exponentially tapered slot}
\newacronym{xlmimo}{XL-MIMO}{extremely large-scale MIMO}
\newacronym{xr}{XR}{extended reality}
\newacronym{z3ro}{Z3RO}{zero third-order distortion}
\newacronym{zed}{ZED}{Zero-Energy device}
\newacronym{zf}{ZF}{zero-forcing}
\newacronym{zmcscg}{ZMCSCG}{zero mean circularly symmetric complex Gaussian}
\newacronym{zmq}{ZMQ}{ZeroMQ}
\newacronym{ztofdm}{ZT-DFT-s-OFDM}{zero-tail DFT-s-OFDM}

%% file: sections/01_introduction.tex
\section{Introduction}
\label{sec:introduction}
\IEEEPARstart{S}{ecuring} wireless communications against eavesdropping has traditionally relied on cryptographic protocols at higher network layers, which face challenges in resource-constrained devices~\cite{poor2017wireless}. \Gls{pls} offers a complementary approach by exploiting the inherent randomness of wireless propagation to provide information-theoretic security guarantees independent of computational assumptions, with theoretical foundations established through Wyner's wiretap channel model~\cite{wyner1975wiretap} and extended to broadcast channels~\cite{csiszar1978broadcast}.
The integration of native security mechanisms at the physical layer has been recognized as a key design principle for future wireless standards~\cite{itu2023imt2030, ara2024pls6g}.
The fundamental limit of \gls{pls} is the secrecy capacity, expressed as $R_s = [R_b - R_e]^+$, where $R_b$ and $R_e$ denote the achievable rates at the legitimate receiver and the eavesdropper, respectively~\cite{liu2009secrecy}, and multiple transmit antennas enable precoding strategies that approach this limit by simultaneously enhancing the legitimate channel while degrading the eavesdropper channel~\cite{khisti2010secure, valliappan2013antenna, nghia2017mimo}.
{{Toward this secrecy capacity limit, three complementary directions have emerged. Wiretap channel codes, including \gls{ldpc}~\cite{klinc2011ldpc}, polar~\cite{wei2016polar}, and lattice
codes~\cite{oggier2016lattice}, are designed to approach the secrecy capacity at the coding
level. Secret key generation exploits wireless channel reciprocity as a separate key-agreement mechanism~\cite{maurer1993secret, rottenberg2021csi}. Multi-antenna precoding adds a spatial dimension by directing signal energy toward legitimate users while suppressing leakage to eavesdroppers, with channel coding serving as an essential complement to approach theoretical capacity limits~\cite{wu2018survey}.}}
{{Building on this spatial dimension, the injection of \gls{an} into the null space of the legitimate channel has proven particularly effective~\cite{goel2008guaranteeing, negi2005secret}, with optimal power allocation studied for fading channels~\cite{zhou2010secure}, cooperative scenarios~\cite{fakoorian2011solutions}, and massive \gls{mimo} deployments~\cite{zhu2016linear, atiya2024cellfree}. Given the extensive literature in this area, the following discussion focuses on works most directly relevant to the proposed approach.}}
\subsection{Background and Motivation}
\label{subsec:pa_gnn}

Although the \gls{pls} techniques discussed above offer strong theoretical foundations, most existing works assume ideal linear hardware. In practice, operating \glspl{pa} close to their saturation point improves energy efficiency, but this regime generates nonlinear distortions whose effects grow with saturation depth~\cite{rapp1991effects, schenk2008rf}. To mitigate these distortions at the signal processing level, distortion-aware precoding has been studied across various architectures, including millimeter-wave systems~\cite{aghdam2019distortion}, massive \gls{mimo}~\cite{aghdam2021distortion}, hybrid beamforming~\cite{moghadam2018energy}, and distributed deployments~\cite{liu2022power}, with fundamental capacity limits characterized in~\cite{bjornson2014hardware, bjornson2019hardware}, where the distortion model in~\cite{bjornson2019hardware} is based on a third-order memoryless nonlinearity, a special case of the higher-order polynomial model adopted in this work.
The \gls{z3ro} precoding framework takes a more active stance, achieving distortion cancellation at receiver locations~\cite{rottenberg2022z3ro, rottenberg2023z3ro}, with {{measurement-based}} validation in~\cite{feys2022measurement}.
While these works treat \gls{pa} nonlinearity as an impairment to be mitigated, a different perspective emerges from the security literature.
From a security perspective, studies have examined how hardware distortions affect secrecy performance in massive \gls{mimo}~\cite{zhu2017secure}, cell-free systems~\cite{zhang2020secure, tahreem2024impact}, and general configurations~\cite{li2024security, wang2024secrecy}. Notably,~\cite{zhu2017secure} observed that additive distortion noise at the transmitter can benefit secrecy when \gls{an} power is insufficient. In contrast,~\cite{zhang2020secure} models hardware impairments as additive Gaussian distortion, showing that transmitter impairments can degrade eavesdropper performance, while user equipment impairments are detrimental to secrecy.

{{Unlike these simplified additive models, realistic \gls{pa} behavior is amplitude-dependent and generates spatially structured distortion that can be directed toward specific locations. This property enables security enhancement by actively exploiting \gls{pa} nonlinearity rather than compensating for it. The proposed framework applies to any {{memoryless}} \gls{pa} that can be characterized by a polynomial model, with the Rapp amplifier used as a representative example for numerical validation.}}
{{Our preliminary work~\cite{ghasemi2025leveraging} demonstrated that Z3RO precoding can leverage \gls{pa} distortion to improve secrecy rates over conventional linear precoders in single-user scenarios.}}
Extending this single-user result to multi-user systems with higher-order \gls{pa} models presents a highly non-convex optimization problem with a nonlinear objective function that traditional methods struggle to solve efficiently.

{{To address this computational challenge, \glspl{gnn} have recently emerged as an effective approach for learning-based precoding. They operate on graph-structured data, naturally capturing antenna-user relationships through
message passing~\cite{scarselli2009graph, hamilton2020graph, battaglia2018relational}, and their permutation equivariance ensures consistent outputs regardless of antenna or user ordering~\cite{zhao2022cnn_gnn}, significantly reducing the
number of learnable parameters compared to fully connected architectures~\cite{shen2021gnn}. In this direction,  \cite{feys2025energy} proposed a \gls{gnn} to maximize sum rate under \gls{pa} nonlinearity,
and~\cite{feys2025quantize} extended this to joint precoding and quantization. Algorithm unfolding provides an interpretable alternative by connecting learned solutions to classical optimization iterations~\cite{chowdhury2021unfolding}, and \gls{ai}-based {{precoder}} has been validated on a distributed \gls{mimo} testbed in~\cite{miao2024testbed}. All these works target rate maximization rather than security, leaving a gap that this work addresses.}}
\subsection{Contributions and Paper Organization}
\label{subsec:contributions}
{{This paper develops a \gls{gnn}-based precoding framework for secure multi-user \gls{miso} communications that deliberately exploits \gls{pa} nonlinearity. The main contributions are as follows:}}
{{
\begin{itemize}
\item A precoding framework is proposed that exploits {{memoryless}} \gls{pa} nonlinearity, modeled by a higher-order polynomial fitted via ridge regression, to direct distortion toward eavesdroppers using only legitimate users' \gls{csi} and without dedicated \gls{an} power allocation.
\item Secrecy rate expressions are derived via Bussgang decomposition for legitimate users and eavesdroppers, and a worst-case lower bound is established for the noiseless eavesdropper; the resulting optimization is highly non-convex, motivating the \gls{gnn}-based solution.
\item A permutation-equivariant \gls{gnn} operating on a bipartite antenna-user graph is designed, achieving parameter efficiency over fully connected architectures and generalizing to varying eavesdropper counts without retraining.
\item The \gls{gnn} is evaluated against \gls{mrt}, \gls{zf}, \gls{an}-aided precoders, and an iterative optimization baseline across \gls{ibo} values of \SI{-1}{\dB} to \SI{-15}{\dB}; at \gls{ibo}~$=$~\SI{-1}{\dB}, it achieves \SI{12.66}{bps/Hz}, outperforming \gls{mrt} by 39.89\%, \gls{zf} by 35.26\%, \gls{an}-aided \gls{mrt} by 17.99\%, and \gls{an}-aided \gls{zf} by 8.67\%, while reducing standard deviation by 58.13--75.31\% across all baselines.
\item A complexity analysis shows that after offline \gls{gpu} training, the \gls{gnn} inference on a \gls{cpu} runs in \SI{0.70}{\milli\second}, well within the \gls{3gpp} \gls{embb} user plane latency budget of \SI{4}{\milli\second}~\cite{3gpp_tr38913}, making it suitable for latency-constrained deployments.
\end{itemize}
}}
The remainder of this paper is organized as follows. \cref{sec:system_model} presents the multi-user \gls{miso} system model with nonlinear \gls{pa} distortion. \cref{sec:sndr_formulation} formulates the secrecy rate optimization problem. \cref{sec:proposed_gnn} presents the proposed \gls{gnn} precoder and its training procedure. \cref{sec:baselines} describes the baseline precoders. \cref{sec:results} evaluates performance across \gls{ibo} levels and assesses robustness to varying eavesdropper counts. \cref{sec:complexity} analyzes the computational complexity and  inference time of all methods. \cref{sec:conclusion} concludes the paper.

\textit{Notation.}
Scalars are denoted by $x$, column vectors by $\mathbf{x}$, and matrices by $\mathbf{X}$.
The $m$-th element of a vector $\mathbf{x}$ is denoted $[\mathbf{x}]_m$. The sets $\mathbb{R}$ and $\mathbb{C}$ denote real and complex numbers, with $\mathbb{R}^{d}$ and $\mathbb{C}^{M \times K}$ denoting $d$-dimensional real vectors and $M \times K$ complex matrices, respectively. Superscripts $(\cdot)^T$, $(\cdot)^H$, and $(\cdot)^*$ indicate transpose, conjugate transpose, and complex conjugate, and $(\cdot)^{\star}$ denotes the optimal value. The Euclidean norm is $\| \cdot \|$, Frobenius norm is $\| \cdot \|_F$, and absolute value is $| \cdot |$. The identity matrix is $\mathbf{I}$. The operator $\odot$ denotes Hadamard product. The function $\mathrm{diag}(\cdot)$ extracts the diagonal of a matrix into a diagonal matrix. The gradient is $\nabla$. Expectation is $\mathbb{E}[\cdot]$, and the positive-part operator is $[x]^+ = \max(0, x)$. Real and imaginary parts are $\mathrm{Re}(\cdot)$ and $\mathrm{Im}(\cdot)$, with $j$ the imaginary unit. Complex Gaussian distributions are $\mathcal{CN}(\boldsymbol{\mu}, \mathbf{C})$ and uniform distributions are $\mathcal{U}(a, b)$. The null space of $\mathbf{H}$ is $\mathrm{null}(\mathbf{H})$, and $\binom{n}{k}$ is the binomial coefficient. Graphs are denoted $\mathcal{G} = (\mathcal{V}, \mathcal{E})$ with vertex set $\mathcal{V}$ and edge set $\mathcal{E}$, $\mathcal{N}(a)$ is the neighborhood of node $a$, and $|\mathcal{N}(a)|$ denotes its cardinality.

%% file: sections/02_system_model.tex
\section{System Model for Multi-User MISO scenario}
\label{sec:system_model}
\begin{figure}[t]
\centering
\includegraphics[width=1\columnwidth]{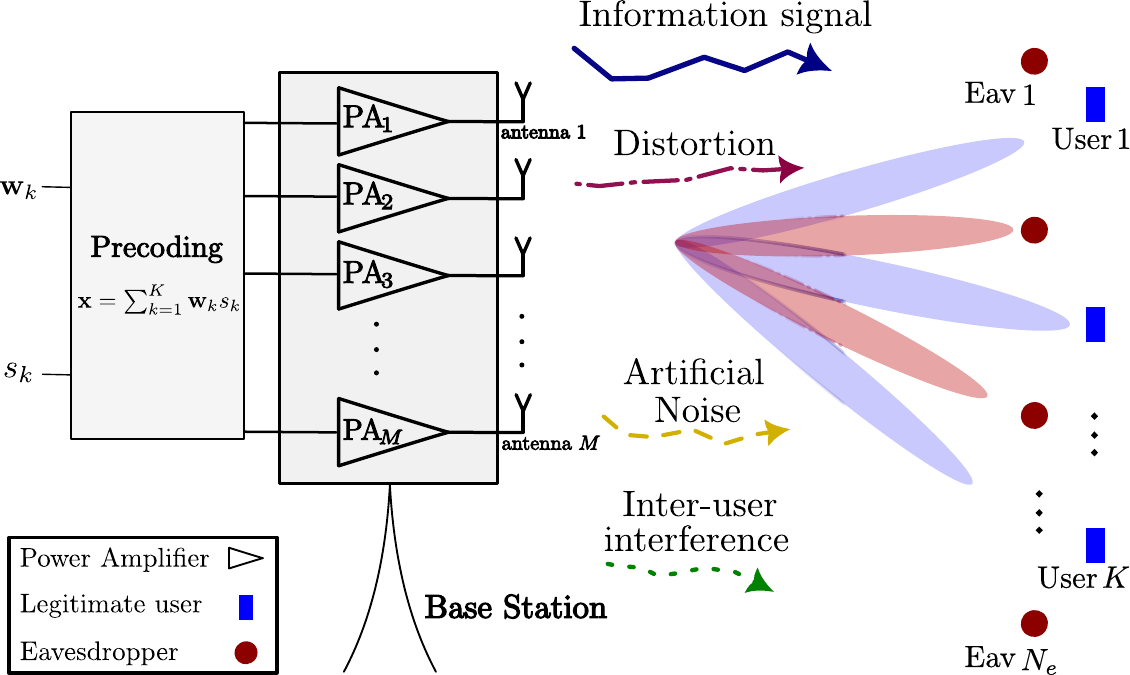}
\caption{Multi-user \gls{miso} wiretap channel system model including transmit antennas with per-antenna nonlinear \glspl{pa}, \( K \) legitimate users, and \( N_e \) eavesdroppers.}
\label{fig:system_model}
\end{figure}

\cref{fig:system_model} illustrates the considered multi-user \gls{miso} wiretap channel system model with nonlinear \gls{pa}. The system consists of a transmitter equipped with \( M \) antennas serving \( K \) legitimate users in the presence of \( N_e \) eavesdroppers. Each antenna is equipped with a nonlinear \gls{pa} that can introduce distortion into the transmitted signals. The \gls{bs} performs precoding, followed by per-antenna nonlinear \gls{pa}s. Signals propagate to \( K \) legitimate users and \( N_e \) eavesdroppers, with information signal, inter-user interference, \gls{pa} distortion, and artificial noise components affecting both receivers.

The channel from the \gls{bs} to the \( k \)-th legitimate user is denoted by \( \vect{h}_k \in \mathbb{C}^{M \times 1} \), and the channel matrix to all legitimate users is \( \mat{H} = [\vect{h}_1, \vect{h}_2, \ldots, \vect{h}_K] \in \mathbb{C}^{M \times K} \). Similarly, the channel from the \gls{bs} to the \( e \)-th eavesdropper is \( \vect{h}_e \in \mathbb{C}^{M \times 1} \), with the eavesdropper channel matrix \( \mat{H}_e \in \mathbb{C}^{M \times N_e} \).
The transmitted signal from \gls{bs} (precoded signal) is given by
\begin{equation}
\label{eq:transmit_signal}
\vect{x} = \sum_{k=1}^{K} \vect{w}_k s_k = \mat{W} \vect{s}
\end{equation}
where \( \vect{x} \in \mathbb{C}^{M \times 1} \) is the precoded signal vector before power amplifiers, \( \mat{W} = [\vect{w}_1, \ldots, \vect{w}_K] \in \mathbb{C}^{M \times K} \) is the precoding matrix, \( \vect{w}_k \in \mathbb{C}^{M \times 1} \) is the precoding vector for user \( k \), \( \vect{s} = [s_1, \ldots, s_K]^T \) contains the unit-variance independent information symbols with \( s_k \sim \mathcal{CN}(0,1) \). The input covariance matrix is \( \mat{C}_x = \mathbb{E}[\vect{x}\vect{x}^H] = \mat{W}\mat{W}^H \) and the total transmit power is limited by \( \norm{\mat{W}}_F^2 = P_t \).

\subsection{Nonlinear Power Amplifier Model}
\label{subsec:pa_model}
Each transmit antenna of the \gls{bs} is equipped with a nonlinear \gls{pa}. The nonlinear behavior of a general \gls{pa} is characterized by the function \( \phi(\cdot) \), which transforms the input signal \( x(t) \) into the output
\begin{equation}
\label{eq:pa_general}
y(t) = \phi(x(t)) = \phi_A(x(t)) e^{j\angle x(t) + \phi_{\varphi}(x(t))}
\end{equation}
where \( \phi_A(\cdot) \) represents the amplitude-to-amplitude (AM/AM) conversion and \( \phi_{\varphi}(\cdot) \) captures the amplitude-to-phase (AM/PM) conversion. This general form covers any memoryless \gls{pa}, and the polynomial approximation and Bussgang decomposition introduced below apply accordingly. For numerical simulations, we use the Rapp model~\cite{rapp1991effects} as a practical example, with AM/AM and AM/PM given by
\begin{equation}      
\label{eq:rapp}              
\phi_A(x_m) = \frac{A|x_m|}{\left(1 + \left|\frac{x_m}{\sqrt{p_{\text{sat}}}}\right|^{2S}\right)^{\!\frac{1}{2S}}}, \quad
\phi_\varphi(x_m) = \frac{|x_m|^q}{1 + \left|\frac{x_m}{B}\right|^q}
\end{equation}
where \( p_{\text{sat}} \) denotes the saturation power of the \gls{pa}, \( x_m \) is the input signal at antenna \( m \), the parameters \( S \) and \( q \) control the transition sharpness between linear and saturated regimes, and coefficients \( A \) and \( B \) determine the AM/PM conversion gain.
The \gls{ibo} characterizes the operating point of the amplifier relative to its saturation level and is defined as \( \text{\gls{ibo} (dB)} = 10\log_{10}(p_{\text{in}} / p_{\text{sat}}) \), where \( p_{\text{in}} = \mathbb{E}[|x_m|^2] \) represents the average input power per antenna. {{The \gls{pa} nonlinearity is modeled by a finite-order polynomial}}; prior works employed lower-order (typically 3rd-order) models, well-justified for mild to moderate saturation where finite polynomial orders provide a well-conditioned fit~\cite{aghdam2019distortion, moghadam2018energy, feys2025energy}.
The \gls{pa} response for complex input \( x_m \) is modeled as
\begin{equation}
\label{eq:pa_polynomial}
\phi(x_m) = \sum_{n=0}^{N} \beta_{2n+1} x_m |x_m|^{2n}
\end{equation}
where \( N \) determines the polynomial order (order \( 2N+1 \)), and \( \{\beta_{2n+1}\} \) are complex coefficients. Only odd-order terms are retained, as even-order terms produce out-of-band spectral regrowth for memoryless \glspl{pa}~\cite{schenk2008rf}.
Coefficients are estimated via \gls{rr} as \(\hat{\boldsymbol{\beta}} = (\mat{X}^H \mat{X} + \alpha \mat{I})^{-1} \mat{X}^H \vect{y}\) where \( \alpha > 0 \) regularizes the ill-conditioned design matrices arising at higher polynomial orders, ensuring numerical stability. The estimated coefficients \( \{\beta_{2n+1}\} \) parameterize the Bussgang decomposition of the \gls{pa} output, which represents the nonlinearly amplified signal as an equivalent linear term plus uncorrelated distortion noise. The polynomial \gls{pa} output can be expressed as~\cite{bussgang1952crosscorrelation}
\begin{equation}
\label{eq:bussgang}
\vect{y} = \mat{G}(\mat{C}_x) \vect{x} + \vect{d}
\end{equation}
where \( \vect{d} \) is the distortion noise vector uncorrelated with \( \vect{x} \) (i.e., \( \mathbb{E}[\vect{d} \vect{x}^H] = \mat{0} \)), and \( \mat{G}(\mat{C}_x) \) is the Bussgang gain matrix computed as~\cite{moghadam2018energy}
\begin{equation}
\label{eq:bussgang_gain}
\mat{G}(\mat{C}_x) = \sum_{n=0}^{N} (n+1)! \beta_{2n+1} \text{diag}(\mat{C}_x)^n.
\end{equation}

The distortion covariance matrix \( \mat{C}_e(\mat{C}_x) = \mathbb{E}[\vect{d}\vect{d}^H] \) characterizes the spatial correlation of the nonlinear distortion across antennas. For circularly symmetric complex Gaussian inputs, these moments can be expressed in closed-form using the input covariance \(\mat{C}_x\), and the resulting distortion covariance can be derived as~\cite{aghdam2021distortion}
\begin{equation}
\label{eq:distortion_cov}
\mat{C}_e(\mat{C}_x) = \sum_{n=1}^{N} \mat{L}_n \left( \mat{C}_x \odot |\mat{C}_x|^{2n} \right) \mat{L}_n^H
\end{equation}
and the coefficient matrices \( \mat{L}_n \) are given by
\begin{equation}
\label{eq:Ln_matrix}
\mat{L}_n = \frac{1}{\sqrt{n+1}} \sum_{l=n}^{N} \binom{l}{n} (l+1)! \beta_{2l+1} \text{diag}(\mat{C}_x)^{l-n}
\end{equation}
where \( \binom{l}{n} \) is the binomial coefficient. Since both \( \mat{G}(\mat{C}_x) \) and \( \mat{C}_e(\mat{C}_x) \) are closed-form polynomial functions of \( \mat{C}_x \), they are differentiable with respect to \( \mat{C}_x \), which is aligned with gradient-based optimization of the precoding matrix.
\subsection{Channel Model}
\label{subsec:channel_models}
For a general channel, \( \vect{h}_k \) can represent any propagation environment; {{Here we consider \gls{los} channels with a \gls{ula} at the transmitter, since the steering-vector structure exposes the angular geometry that our security mechanism exploits, making the analysis transparent.}} The channel vector to user \( k \) is  $\mathbf{h}_k = \sqrt{g_k},\mathbf{a}(\theta_k)$ 
where \( g_k \) is the path loss coefficient, {{\( \theta_k \in [0,\pi]\) is the angle-of-arrival (AoA) for user \( k\), and \( \vect{a}(\theta_k)\) is the array response vector of the ULA}}. For a \gls{ula} with half-wavelength antenna spacing \( d = \lambda_c/2 \), the array response is $[\mathbf{a}(\theta_k)]_m = \exp(-j\pi \cos(\theta_k),m)$
where \( \lambda_c \) is the carrier wavelength, and \( m = 0, 1, \ldots, M-1 \). For \gls{pls} analysis, we model \( N_e \) eavesdropper channels to evaluate secrecy rates. The base angles are uniformly spaced as \( \theta_{e,i}^{b} = (i-1) \pi / (N_e-1) \) for \( i = 1, \ldots, N_e \). A random shift \( \delta_s \sim \mathcal{U}(0, \pi/(N_e-1)) \) is applied to all eavesdroppers, yielding final angles \( \theta_{e,i} = \theta_{e,i}^{b} + \delta_s \). The shift range \( \pi/(N_e-1) \) equals the inter-eavesdropper spacing, so different realizations cover distinct eavesdropper positions across the angular range. {{Eavesdropper and user angles can be arbitrarily close, with no minimum separation enforced. This setup evaluates the precoder against the full range of eavesdropper positions, including those near legitimate users.}}

%% file: sections/03_sndr_formulation.tex
\section{Problem Formulation}
\label{sec:sndr_formulation}
{{In this section, we derive the \gls{snidr} at legitimate users and eavesdroppers using the Bussgang decomposition in \cref{eq:bussgang}. \Cref{subsec:secrecy_rate_lower_bound} establishes a worst-case lower bound on the secrecy rate, and \cref{subsec:secrecy_rate} formulates the resulting optimization problem.}}

The received signal at legitimate user \( k \) is
\begin{equation}
\label{eq:received_signal_user}
y_k = \vect{h}_k^H (\mat{G}(\mat{C}_x) \vect{x} + \vect{d}) + \nu_k
\end{equation} where \( \vect{h}_k \in \mathbb{C}^{M \times 1} \) is the channel vector from the transmitter to user \( k \), and \( \nu_k \sim \mathcal{CN}(0, \sigma_{\nu}^2 ) \) is \gls{iid} \gls{awgn}. Substituting \( \vect{x} = \mat{W}\vect{s} \) from \cref{eq:transmit_signal} and expanding, we obtain
\begin{equation}
\label{eq:received_signal_user_decomp}
y_k = \underbrace{\vect{h}_k^H \mat{G}(\mat{C}_x) \vect{w}_k s_k}_{\text{Desired Signal}}
+ \underbrace{\sum_{j \neq k} \vect{h}_k^H \mat{G}(\mat{C}_x) \vect{w}_j s_j}_{\text{Inter-User Interference}} + \underbrace{\vect{h}_k^H \vect{d}}_{\text{Distortion}} + \underbrace{\nu_k}_{\text{Noise}}
\end{equation}
{{where the four terms represent the desired signal for user \( k \), inter-user interference, \gls{pa} distortion, and thermal noise, respectively. The desired signal and interference are both affected by the Bussgang gain \( \mat{G}(\mat{C}_x) \), which captures the amplitude compression due to \gls{pa} nonlinearity. Note that \( \vect{d} \) also includes nonlinear inter-user mixing, since the \gls{pa} acts on the combined signal of all users; this is captured within the distortion term. Unlike heuristic additive distortion models, the distortion term \( \vect{h}_k^H \vect{d} \) propagates through the channel with spatial correlation characterized by \( \mat{C}_e(\mat{C}_x) \),
enabling deliberate exploitation through precoding design.}}

The received signal at eavesdropper \( e \) is
\begin{equation}
\label{eq:received_signal_eve}
y_e = \vect{h}_e^H (\mat{G}(\mat{C}_x) \vect{x} + \vect{d}) + \varepsilon_e
\end{equation}
where \( \vect{h}_e \in \mathbb{C}^{M \times 1} \) is the channel vector from the transmitter to eavesdropper \( e \) and \( \varepsilon_e \sim \mathcal{CN}(0, \sigma_{\varepsilon}^2) \) is \gls{iid} \gls{awgn}.

The eavesdropper receives a superposition of all \( K \) users' signals. When attempting to decode user \( k \)'s signal, the eavesdropper treats \( s_k \) as the desired signal and \( \{s_j\}_{j \neq k} \) as interference
\begin{equation}
\label{eq:received_signal_eve_decomp}
y_e = \underbrace{\vect{h}_e^H \mat{G}(\mat{C}_x) \vect{w}_k s_k}_{\text{Desired Signal (User } k\text{)}}
+ \underbrace{\sum_{j \neq k} \vect{h}_e^H \mat{G}(\mat{C}_x) \vect{w}_j s_j}_{\text{Inter-User Interference}}
+ \underbrace{\vect{h}_e^H \vect{d}}_{\text{Distortion}} + \underbrace{\varepsilon_e}_{\text{Noise}}
\end{equation}
where each term corresponds to the signal component from user \( k \), interference from other users, \gls{pa} distortion, and thermal noise, respectively. The eavesdropper experiences the same
Bussgang gain \( \mat{G}(\mat{C}_x) \) and distortion covariance \( \mat{C}_e(\mat{C}_x) \) as legitimate users, since these depend only on the transmit signal statistics \( \mat{C}_x = \mat{W}\mat{W}^H \), not on the receive channel.
{{This treats other users' signals as interference; the lower bound in \cref{subsec:secrecy_rate_lower_bound} considers an eavesdropper that cancels them perfectly, yielding a more conservative security metric.}}

The \gls{snidr} at legitimate user \( k \) is the ratio of desired signal power to the total interference-plus-noise-plus-distortion power
\begin{equation}
\label{eq:sindr_user}
\text{SNIDR}_k^{\text{legit}} = \frac{|\vect{h}_k^H \mat{G}(\mat{C}_x) \vect{w}_k|^2}{\sum_{j \neq k} |\vect{h}_k^H \mat{G}(\mat{C}_x) \vect{w}_j|^2 + \vect{h}_k^H \mat{C}_e(\mat{C}_x) \vect{h}_k + \sigma_{\nu}^2}.
\end{equation}
The distortion power in the denominator depends on both the channel \( \vect{h}_k \) and the distortion covariance \( \mat{C}_e(\mat{C}_x) \), which is a function of the input power distribution across antennas.

The \gls{snidr} at eavesdropper \( e \) when attempting to decode user \( k \)'s signal is
\begin{equation}
\label{eq:sindr_eve}
\text{SNIDR}_{k,e}^{\text{eav}} = \frac{|\vect{h}_e^H \mat{G}(\mat{C}_x) \vect{w}_k|^2}{\sum_{j \neq k} |\vect{h}_e^H \mat{G}(\mat{C}_x) \vect{w}_j|^2 + \vect{h}_e^H \mat{C}_e(\mat{C}_x) \vect{h}_e + \sigma_{\varepsilon}^2}
\end{equation}
where the notation \( \text{SNIDR}_{k,e}^{\text{eav}} \) indicates that eavesdropper \( e \) is targeting user \( k \)'s signal. This shows the selective eavesdropping scenario that each eavesdropper needs to choose which user to decode from the received superposition.
To further degrade eavesdropper performance, the precoding matrix can be extended to include \gls{an} vectors designed to lie in the null space of legitimate user channels~\cite{goel2008guaranteeing}
\begin{equation}
\label{eq:extended_precoding}
\mat{W}_{\text{ext}} = [\vect{w}_1, \ldots, \vect{w}_K, \vect{z}_1, \ldots, \vect{z}_{M-K}] \in \mathbb{C}^{M \times M}
\end{equation}
where \( \mat{Z} = [\vect{z}_1, \ldots, \vect{z}_{M-K}] \) satisfies the null space constraint \( \mat{H}^H \mat{Z} = \mat{0} \), ensuring that \gls{an} does not interfere with legitimate users in linear systems. The extended transmit signal becomes \( \vect{x} = \sum_{k=1}^{K}
\vect{w}_k s_k + \sum_{j=1}^{M-K} \vect{z}_j v_j \), where \( v_j \sim \mathcal{CN}(0, 1) \) are independent \gls{an} symbols, and the input covariance is
\begin{equation}
\mat{C}_x = \mat{W}_{\text{ext}} \mat{W}_{\text{ext}}^H = \sum_{k=1}^{K} \vect{w}_k \vect{w}_k^H + \sum_{j=1}^{M-K} \vect{z}_j \vect{z}_j^H.
\end{equation}
However, \gls{pa} nonlinearity breaks the null space orthogonality. Under Bussgang decomposition, the received signal at user \( k \) includes an \gls{an} leakage term
\begin{align}
y_k &= \vect{h}_k^H \mat{G}(\mat{C}_x) \vect{w}_k s_k + \sum_{j \neq k} \vect{h}_k^H \mat{G}(\mat{C}_x) \vect{w}_j s_j \nonumber \\
&\quad + \sum_{j=1}^{M-K} \vect{h}_k^H \mat{G}(\mat{C}_x) \vect{z}_j v_j + \vect{h}_k^H \vect{d} + \nu_k,
\end{align}
comprising signal, inter-user interference, \gls{an} leakage, distortion, and noise terms, respectively. The \gls{an} leakage term \( \sum_{j} \vect{h}_k^H \mat{G}(\mat{C}_x) \vect{z}_j v_j \neq 0 \) even though \( \vect{h}_k^H \mat{Z} = \vect{0}^T \), since the Bussgang gain \( \mat{G}(\mat{C}_x) \) depends nonlinearly on \( \mat{C}_x = \mat{W}_{\text{ext}} \mat{W}_{\text{ext}}^H \). {{In the linear regime (low \gls{ibo}), \( \mat{G}(\mat{C}_x) \) reduces to a scaled identity \( \beta_1 \mat{I} \), so \( \vect{h}_k^H \mat{G}(\mat{C}_x) \mat{Z} = \beta_1\,\vect{h}_k^H \mat{Z} = \vect{0}^T \) and the null-space orthogonality is preserved. As the \gls{pa} approaches saturation, \( \mat{G}(\mat{C}_x) \) is no longer proportional to the identity and \gls{an} leakage at legitimate users emerges.}} The \gls{snidar} at legitimate user \( k \) in the presence of \gls{an} is given in \cref{eq:snidar_user_an}, where the denominator includes inter-user interference, \gls{an} leakage power, distortion power, and noise, respectively.
\begin{align}
\label{eq:snidar_user_an}
&\text{SNIDAR}_k^{\text{legit}} = |\vect{h}_k^H \mat{G}(\mat{C}_x) \vect{w}_k|^2 \nonumber \\
&\times \!\bigg(\!\sum_{j \neq k} |\vect{h}_k^H \mat{G}(\mat{C}_x) \vect{w}_j|^2 + \!\sum_{j=1}^{M-K} |\vect{h}_k^H \mat{G}(\mat{C}_x) \vect{z}_j|^2 \nonumber \\
&+ \vect{h}_k^H \mat{C}_e(\mat{C}_x) \vect{h}_k + \sigma_{\nu}^2 \bigg)^{\!-1}
\end{align}
Similarly, eavesdropper \( e \) receives \gls{an} leakage through the Bussgang gain, yielding
\begin{align}      
\label{eq:snidar_eve_an}
&\text{SNIDAR}_{k,e}^{\text{eav}} = |\vect{h}_e^H \mat{G}(\mat{C}_x) \vect{w}_k|^2 \nonumber \\
&\times \!\bigg(\!\sum_{j \neq k} |\vect{h}_e^H \mat{G}(\mat{C}_x) \vect{w}_j|^2 + \!\sum_{j=1}^{M-K} |\vect{h}_e^H \mat{G}(\mat{C}_x) \vect{z}_j|^2 \nonumber \\
&+ \vect{h}_e^H \mat{C}_e(\mat{C}_x) \vect{h}_e + \sigma_{\varepsilon}^2 \bigg)^{\!-1}
\end{align} 
Since eavesdropper channels lie outside the null space, they experience substantially larger \gls{an} power, preserving the security benefit despite leakage at legitimate users.
\subsection{Lower Bound on Secrecy Rate}
\label{subsec:secrecy_rate_lower_bound}

To characterize the achievable performance under worst-case eavesdropping scenarios, we derive a lower bound on the secrecy rate following the approach of~\cite{zhou2010secure}. {{This bound will be a benchmark for the worst-case secrecy performance.}}

By considering the worst-case scenario where eavesdropper \( e \) is assumed to be noiseless (\( \sigma_{\varepsilon}^2 \to 0 \)) and capable of perfectly canceling inter-user interference, the eavesdropper \gls{sdar} when targeting user \( k \)'s signal becomes
\begin{equation}
\label{eq:sdar_noiseless}
\text{SDAR}_{k,e}^{\text{noiseless}} = \frac{|\vect{h}_e^H \mat{G}(\mat{C}_x) \vect{w}_k|^2}{\sum_{j=1}^{M-K} |\vect{h}_e^H \mat{G}(\mat{C}_x) \vect{z}_j|^2 + \vect{h}_e^H \mat{C}_e(\mat{C}_x) \vect{h}_e}.
\end{equation}
The numerator represents the desired signal power for user \( k \), while the denominator comprises
\gls{an} leakage and \gls{pa} distortion powers experienced by the eavesdropper. This assumption provides an upper bound on eavesdropper capability, as any practical eavesdropper with noise or residual interference performs worse.
The worst-case scenario considers the eavesdropper with the highest \gls{sdar} among all \( N_e \) eavesdroppers, \( \text{SDAR}_k^{\text{eav,worst}} = \max_{e=1,\ldots,N_e} \text{SDAR}_{k,e}^{\text{noiseless}} \). This represents the strongest potential eavesdropper for user \( k \)'s transmission, providing a conservative estimate for security analysis.

Using the worst-case eavesdropper \gls{sdar}, we establish an achievable lower bound on the sum secrecy rate given in \cref{eq:secrecy_rate_lower_bound}, where \( \text{SNIDR}_k^{\text{legit}} \)
is given in \cref{eq:sindr_user}
\begin{align}
\label{eq:secrecy_rate_lower_bound}
R_{\text{s}}^{\text{lower}} = \sum_{k=1}^{K} \big[&\log_2(1 + \text{SNIDR}_k^{\text{legit}}) \nonumber \\
&- \log_2(1 + \text{SDAR}_k^{\text{eav,worst}})\big]^+  
\end{align}
This bound represents a more challenging security scenario {{than the secrecy rate computed from $\text{SNIDR}_{k,e}^{\text{eav}}$ in \cref{eq:sindr_eve},}} as it assumes the eavesdropper can identify and decode the strongest user signal without interference or noise. {{This lower bound~\cite{zhou2010secure} provides security guarantees that hold against the most capable eavesdropper.}}
\subsection{Secrecy Rate and Optimization Problem}
\label{subsec:secrecy_rate}
For notational convenience, let \( \gamma_k \triangleq \text{SNIDR}_k^{\text{legit}} \) and \( \gamma_{e,k} \triangleq \text{SNIDR}_{k,e}^{\text{eav}} \). The achievable rates are \( R_k = \log_2(1 + \gamma_k) \) and \( R_{e,k} = \log_2(1 + \gamma_{e,k}) \), and the secrecy rate for user \( k \) is \( R_{s,k} = [R_k - R_{e,k}]^+ \)~\cite{wyner1975wiretap}, where \( [x]^+ = \max(0, x) \). The sum secrecy rate is \( R_s = \sum_{k=1}^{K} R_{s,k} \). As the eavesdropper channels are unknown to the transmitter, the optimization objective is the expected sum secrecy rate over the eavesdropper channel distribution
\begin{equation}
\label{eq:optimization_problem}
\begin{aligned}
& \underset{\mat{W}}{\text{maximize}} \quad \mathbb{E}_{\mat{H}_e}\left[R_s\right] \\
& \text{subject to} \quad \norm{\mat{W}}_F^2 \leq P_t.
\end{aligned}
\end{equation}
{{Due to the \gls{pa} nonlinearity, the Bussgang gain and distortion covariance are polynomial in the input covariance, adding to the inherent non-convexity from the rate-difference structure and inter-user interference, with many local maxima that classical convex methods cannot handle. While the worst-case lower bound is useful for evaluating security guarantees, using it as the training objective gives a less smooth landscape and overfits to specific angular positions; the expected sum secrecy rate is therefore used for training.}}
A conventional approach to this problem is projected gradient ascent~\cite{boyd2004convex}, where at iteration \( t \) the precoding matrix is updated as
\begin{equation}
\label{eq:gradient_ascent}
\mat{W}^{(t+1)} = \mathcal{P}_{P_t}\left(\mat{W}^{(t)} + \mu^{(t)} \nabla_{\mat{W}} \mathbb{E}_{\mat{H}_e}[R_{\text{s}}]\right)
\end{equation}
where \( \mu^{(t)} \) is the step size, \( \nabla_{\mat{W}} \) denotes the complex gradient with respect to \( \mat{W} \), and \( \mathcal{P}_{P_t}(\cdot) \) projects onto the power constraint
\begin{equation}
\label{eq:power_projection}
\mathcal{P}_{P_t}(\mat{W}) = \begin{cases}
\mat{W} & \text{if } \norm{\mat{W}}_F^2 \leq P_t \\
\sqrt{{P_t}/{\norm{\mat{W}}_F^2}}\, \mat{W} & \text{otherwise}
\end{cases}.
\end{equation}
The step size \( \mu^{(t)} = \mu_0 \cdot \rho^t \) is decayed exponentially with \( \rho < 1 \) to balance exploration and convergence. However, gradient ascent requires hundreds to thousands of iterations per channel realization and is sensitive to initialization, making it impractical within the channel coherence time.

To address this, a \gls{gnn}-based approach is proposed in \cref{sec:proposed_gnn} that learns to approximate the optimal precoder from \( \mat{H} \) in a single forward pass~\cite{shen2021gnn, chowdhury2021unfolding, feys2025quantize}. {{ While \glspl{mlp} are universal approximators~\cite{hornik1989universal}, they must learn permutation equivariance from data, and \glspl{cnn} suit grid-structured inputs. In contrast, \glspl{gnn} model the channel as a bipartite antenna-user graph, where permuting antennas or users also permutes the output precoder~\cite{zhao2022cnn_gnn, battaglia2018relational}.}} {{This built-in equivariance decouples the parameter count from \( M \) and \( K \). At matched depth and width of six layers and \( 64 \) features per layer, the \gls{gnn} uses \( 50{,}000 \) parameters versus \( 450{,}000 \) for the \gls{ccnn} and \( 270{,}000{,}000 \) for the \gls{mlp}~\cite{feys2025energy}.}} Building on recent \gls{gnn} applications to hardware nonlinearities~\cite{feys2025energy, feys2025quantize}, the \gls{gnn} exploits the spatial structure of \gls{pa} distortion to direct it toward eavesdroppers while preserving signal quality at legitimate users.


%% file: sections/04_proposed_gnn.tex
\section{Proposed GNN-Based Precoding}
\label{sec:proposed_gnn}
The proposed \gls{gnn}-based precoding architecture learns to map channel matrices directly to precoding matrices while implicitly exploiting \gls{pa} distortion for security enhancement. The \gls{gnn} operates using only legitimate user channel information \( \mat{H} \) as input, without any knowledge of eavesdropper \gls{csi}, making the precoder robust to unknown eavesdropper locations without requiring eavesdropper channels.


\subsection{Graph Representation}
\label{subsec:graph_representation}
The precoding problem can be modeled as a bipartite graph \( \mathcal{G} = (\mathcal{V}_m \cup \mathcal{V}_k, \mathcal{E}) \), where \( \mathcal{V}_m = \{1, 2, \ldots, M\} \) represents the set of \( M \) transmit antennas, \( \mathcal{V}_k = \{1, 2, \ldots, K\} \) represents the set of \( K \) users, and \( \mathcal{E} \subseteq \mathcal{V}_m \times \mathcal{V}_k \) represents edges between antennas and users. Each edge \( (m, k) \in \mathcal{E} \) is associated with the channel coefficient \( h_{m,k} \), which represents the complex channel gain from antenna \( m \) to user \( k \). Since neural networks typically operate on real-valued features, each channel coefficient can be decomposed into its real and imaginary parts, yielding a 2-dimensional edge feature \( \vect{h}_{m,k} = [\text{Re}(h_{m,k}), \text{Im}(h_{m,k})]^T \in \mathbb{R}^2 \). This bipartite graph structure captures the antenna-user connectivity in \gls{miso} systems, where the precoding weight \( w_{m,k} \) (the complex weight from antenna \( m \) to user \( k \)) corresponds to the edge \( (m, k) \) in the graph. The \gls{gnn} processes this graph structure through iterative message passing to compute optimal precoding weights.

\subsection{Message Passing}
\label{subsec:message_passing}
\begin{figure}[t]
\centering
\includegraphics[width=0.8\columnwidth]{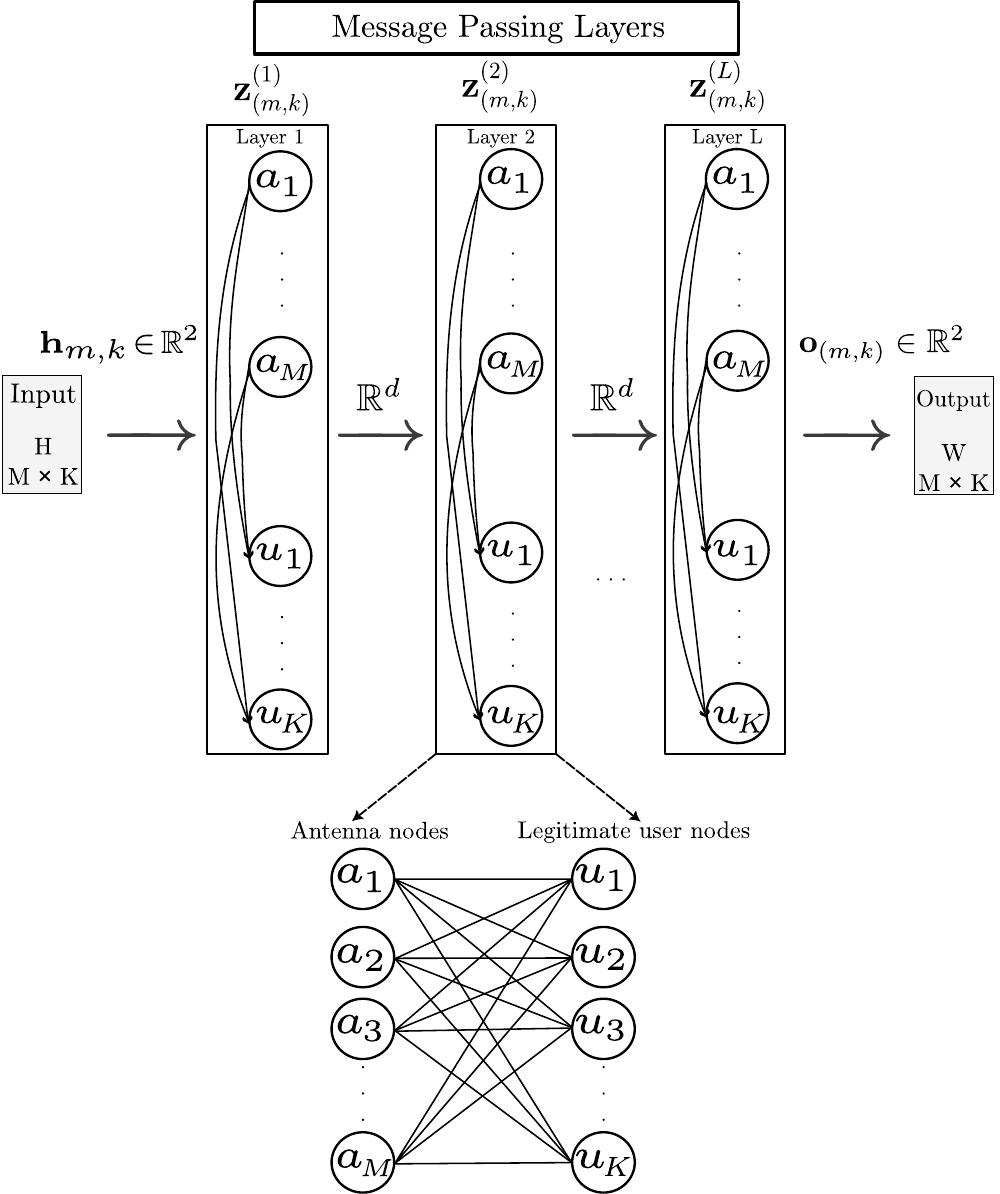}
\caption{{GNN architecture with \( L \) message passing layers on a bipartite graph between antenna nodes \( a_1, \ldots, a_M \) and user nodes \( u_1, \ldots, u_K \). Edges carry the channel input \( \vect{h}_{m,k} \in \mathbb{R}^2 \), the hidden representation \( \vect{z}_{(m,k)}^{(\ell)} \in \mathbb{R}^d \), and the precoding output \( \vect{o}_{(m,k)} \in \mathbb{R}^2 \), while node features are formed by mean-pooling neighboring edge features.}}
\label{fig:gnn_architecture}
\end{figure}

The \gls{gnn} architecture consists of an input layer, \( L \) hidden layers, and an output layer, as illustrated in \cref{fig:gnn_architecture}.
Each layer updates edge representations through aggregation and learnable transformations, where \( \vect{z}_{(m,k)}^{(\ell)} \in \mathbb{R}^d \) denotes the feature vector of edge \( (m,k) \) at layer \( \ell \), with \( d \) being the hidden dimension. The input layer transforms the 2-dimensional channel features to the hidden dimension via \(\vect{z}_{(m,k)}^{(0)} = \mat{W}^{(0)}\vect{h}_{m,k} \in \mathbb{R}^d\), where \(\vect{h}_{m,k} = [\mathrm{Re}(h_{m,k}), \mathrm{Im}(h_{m,k})]^T \in \mathbb{R}^2\) and \(\mat{W}^{(0)} \in \mathbb{R}^{d \times 2}\) is a learnable projection. At each hidden layer \( \ell = 1, \ldots, L \), the edge feature \( \vect{z}_{(m,k)}^{(\ell)} \) is computed by combining three components: the current edge representation \(\vect{z}_{(m,k)}^{(\ell-1)}\), aggregated features from neighboring edges at the antenna endpoint \(\bar{\vect{z}}_m^{(\ell-1)}\), and aggregated features at the user endpoint \(\bar{\vect{z}}_k^{(\ell-1)}\). The layer update is
{{\begin{equation}
\label{eq:edge_update}
\vect{z}_{(m,k)}^{(\ell)} = \sigma\!\left( \mat{W}_{\text{e}}^{(\ell)} \vect{z}_{(m,k)}^{(\ell-1)} + \mat{W}_{\text{a}}^{(\ell)} \bar{\vect{z}}_m^{(\ell-1)} + \mat{W}_{\text{u}}^{(\ell)} \bar{\vect{z}}_k^{(\ell-1)} \right)
\end{equation}}}
where \( \sigma(\cdot) \) denotes the leaky rectified linear unit (LeakyReLU) activation function and \( \mat{W}_{\text{e}}^{(\ell)}, \mat{W}_{\text{a}}^{(\ell)}, \mat{W}_{\text{u}}^{(\ell)} \in \mathbb{R}^{d \times d} \) are learnable weight matrices for edges, antennas, and users, respectively. The aggregated neighborhood features are computed using mean pooling
\begin{equation}
\label{eq:neighbor_mean}
\bar{\vect{z}}_i^{(\ell-1)} = \frac{1}{|\mathcal{N}(i)|} \sum_{j \in \mathcal{N}(i)} \vect{z}_{(i,j)}^{(\ell-1)}, \quad i \in \{m, k\}.
\end{equation}
This mean aggregation enables each edge to incorporate contextual information from all edges sharing either endpoint, capturing both per-antenna and per-user signal characteristics while maintaining scale invariance.
\subsection{Output Layer and Normalization}
\label{subsec:output_layer}
The output layer follows the same structure as \cref{eq:edge_update} with linear activation
\begin{equation}
\label{eq:output_layer}
\vect{o}_{(m,k)} = \mat{W}_{\text{e}}^{\text{out}} \vect{z}_{(m,k)}^{(L)} + \mat{W}_{\text{a}}^{\text{out}} \bar{\vect{z}}_m^{(L)} + \mat{W}_{\text{u}}^{\text{out}} \bar{\vect{z}}_k^{(L)}
\end{equation}
where \( \mat{W}_{\text{e}}^{\text{out}}, \mat{W}_{\text{a}}^{\text{out}}, \mat{W}_{\text{u}}^{\text{out}} \in \mathbb{R}^{2 \times d} \) and \( \vect{o}_{(m,k)} = [\mathrm{Re}(\tilde{w}_{m,k}), \mathrm{Im}(\tilde{w}_{m,k})]^T \) forms the raw precoding matrix \( \widetilde{\mat{W}} \in \mathbb{C}^{M \times K} \). To satisfy the power constraint \( \norm{\mat{W}}_F^2 = P_t \), normalization is applied $\mathbf{W} = \sqrt{P_t},\widetilde{\mathbf{W}}/|\widetilde{\mathbf{W}}|_F$.
\subsection{Loss Function and Training Procedure}
\label{subsec:training_objective}
\label{subsec:training_procedure}
The training objective is to maximize the expected sum secrecy rate defined in \cref{eq:optimization_problem}, formulated as minimizing the loss function \( \mathcal{L}(\boldsymbol{\varphi}) = -\mathbb{E}_{\mat{H} \sim \mathcal{D}} \left[ \mathbb{E}_{\mat{H}_e}\left[R_s\right] \right], \)
where \( \boldsymbol{\varphi} \) denotes all learnable parameters and \( \mathcal{D} \) is the distribution of legitimate user channel matrices. The sum secrecy rate is computed using the Bussgang gain \( \mat{G}(\mat{C}_x) \) and distortion covariance \( \mat{C}_e(\mat{C}_x) \) from \cref{eq:bussgang_gain,eq:distortion_cov} with \( \mat{C}_x = \mat{W}\mat{W}^H \). The complete training procedure is summarized in \cref{alg:gnn_training}, where the outer expectation is approximated by averaging over minibatches of \( B \) legitimate channel realizations with parameters updated via the Adam optimizer, and the inner expectation is approximated by averaging over \( N_e \) eavesdropper realizations.
\begin{algorithm}[t]
\caption{Distortion-Aware GNN-Based PLS Sum Secrecy Rate Maximization}
\label{alg:gnn_training}
\begin{algorithmic}[1]
\State \textbf{Input:} Training set \( \mathcal{D}_{\text{train}} \), validation set \( \mathcal{D}_{\text{val}} \), learning rate \( \eta \), batch size \( B \), eavesdropper samples \( N_e \)
\State \textbf{Output:} Optimized parameters \( \boldsymbol{\varphi}^* \)
\State Initialize \( \boldsymbol{\varphi} \) randomly, \( \mathcal{L}^* \leftarrow \infty \)
\For{\( t = 1, 2, \ldots, T \)}
    \For{each minibatch \( \{\mat{H}^{(b)}\}_{b=1}^{B} \subset \mathcal{D}_{\text{train}} \)}
        \State \( \mat{W}^{(b)} \leftarrow f_{\boldsymbol{\varphi}}(\mat{H}^{(b)}) \)
        \State \( \mathcal{L} \leftarrow -\frac{1}{B} \sum_{b=1}^{B} \frac{1}{N_e} \sum_{e=1}^{N_e} R_s^{(b,e)} \)
        \State \( \boldsymbol{\varphi} \leftarrow \boldsymbol{\varphi} - \eta \nabla_{\boldsymbol{\varphi}} \mathcal{L} \)
    \EndFor
    \State \( \mathcal{L}_{\text{val}} \leftarrow \) Evaluate \( \mathcal{L} \) on \( \mathcal{D}_{\text{val}} \)
    \If{\( \mathcal{L}_{\text{val}} < \mathcal{L}^* \)}
        \State \( \boldsymbol{\varphi}^* \leftarrow \boldsymbol{\varphi} \), \( \mathcal{L}^* \leftarrow \mathcal{L}_{\text{val}} \)
    \EndIf
\EndFor
\State \Return \( \boldsymbol{\varphi}^* \)
\end{algorithmic}
\end{algorithm}
The dataset consists of channel matrices generated according to the \gls{los} model in \cref{subsec:channel_models} with uniformly random user angles, partitioned into \( N_{\text{train}} = 350{,}000 \) training, \( N_{\text{val}} = 25{,}000 \) validation, and \( N_{\text{test}} = 16{,}000 \) test samples. The model is trained using the Adam optimizer~\cite{kingma2014adam} with cosine learning rate annealing (\( \eta = 5\times10^{-3} \), \( \eta_{\min} = 10^{-4} \), batch size \( B = 128 \)). By construction, the edge-based message passing is permutation equivariant with respect to both antennas and users~\cite{zhao2022cnn_gnn}, ensuring consistent precoding outputs regardless of their ordering in \( \mat{H} \).

%% file: sections/05_baseline_methods.tex
\section{Baseline Methods}
\label{sec:baselines}
To evaluate the performance of the proposed \gls{gnn}-based precoding, we compare against several baseline methods ranging from classical precoders to AN-based precoders and computationally intensive iterative optimization. All baselines use the same \gls{pa} model and Bussgang decomposition for fair comparison.

\subsection{Classical Precoders: MRT, ZF}
\label{subsec:classical_precoders}
\gls{mrt} maximizes received signal power by aligning the precoding vector for user \( k \) with its channel conjugate as \( \vect{w}_k = \vect{h}_k^* / \norm{\vect{h}_k} \), followed by power normalization to satisfy the total power constraint. {{While computationally simple, \gls{mrt} is designed to maximize received signal power per user, which is optimal in the single-user case without eavesdroppers. In the multi-user secrecy setting considered here, it serves as a reference baseline rather than a secrecy-aware precoder.}} {{Appendix~\ref{app:appendix_loss_derivation} provides intuition for this through a gradient analysis under the simplified single-user linear \gls{pa} regime.}}
\gls{zf} precoding eliminates inter-user interference by computing the pseudo-inverse \( \mat{W}_{\text{ZF}} = \mat{H} (\mat{H}^H \mat{H})^{-1} \) followed by power normalization, requiring \( M \geq K \) for the matrix inversion to exist. Although \gls{zf} improves \gls{snidr} in multi-user scenarios by eliminating interference, it does not consider eavesdropper channels and may suffer from noise amplification when users have similar channel directions.

\subsection{Artificial Noise Precoders: AN-aided MRT, AN-aided ZF}
\label{subsec:artificial_noise}

\gls{an}~\cite{goel2008guaranteeing} injects jamming into the null space of the legitimate channel to degrade eavesdroppers without affecting users. The transmitted signal is modeled as $\vect{x} = \mat{W}_{\text{ext}} \vect{s}_{\text{ext}}$, where $\vect{s}_{\text{ext}} = [s_1, \ldots, s_K, v_1, \ldots, v_{M-K}]^T \in \mathbb{C}^{M}$ stacks the information symbols $s_k$ and \gls{an} symbols $v_j \sim \mathcal{CN}(0,1)$. The extended precoding matrix is
\begin{equation}
\label{eq:an_precoder}
\mat{W}_{\text{ext}} = \Bigl[\sqrt{\alpha P_t}\cdot \mat{W}_{\text{norm}},\, \sqrt{(1-\alpha) P_t}\cdot \mat{Z}_{\text{norm}}\Bigr] \in \mathbb{C}^{M \times M}
\end{equation}
where $\mat{W}_{\text{norm}} \in \mathbb{C}^{M \times K}$ is the unit-norm information precoder (\gls{mrt} or \gls{zf}) and $\mat{Z}_{\text{norm}} \in \mathbb{C}^{M \times (M-K)}$ is the unit-norm \gls{an} precoder constructed via \gls{svd} of $\mat{H}^H$, satisfying $\mat{H}^H \mat{Z} = \mathbf{0}_{K \times (M-K)}$, with $\|\mat{W}_{\text{ext}}\|_F^2 = P_t$, requiring $M > K$. The power split $\alpha$ is selected via grid search to maximize the mean sum secrecy rate. With $\mat{W}_{\text{norm}}$ set to \gls{mrt} or \gls{zf}, this yields the AN-aided MRT and AN-aided ZF baselines.
{{However, \gls{pa} nonlinearity breaks the null-space assumption underlying this approach. Under \gls{pa} nonlinearity, different antennas carry different instantaneous signal powers and therefore receive different gains from their respective \glspl{pa}, described by the input-dependent diagonal gain matrix \( \mat{G}(\mat{C}_x) \).}}

\subsection{Gradient-Based Optimization}
\label{subsec:optimization}
{{As an optimization-based baseline, denoted Opt-GNN, we implement direct gradient ascent of the precoding matrix \( \mat{W} \) to maximize the sum secrecy rate, as formulated in \cref{eq:optimization_problem}. This optimization scheme is initialized with the \gls{gnn} output as the best performing approach, then iteratively refines via gradient ascent with projection onto the power constraint. Each iteration requires computing \gls{snidr} for all users and eavesdroppers, plus gradient computation.}}

%% file: sections/06_numerical_results.tex
\section{Numerical Results}
\label{sec:results}
In this section, we present the numerical results to evaluate the performance of the proposed \gls{gnn}-based precoding against the baseline methods described in \cref{sec:baselines}.

\subsection{Simulation Setup and Analytical Validation}
\label{subsec:simulation_setup}
\label{subsec:sdar_analysis}
\begin{figure}[t]
\centering
\includegraphics[width=\columnwidth]{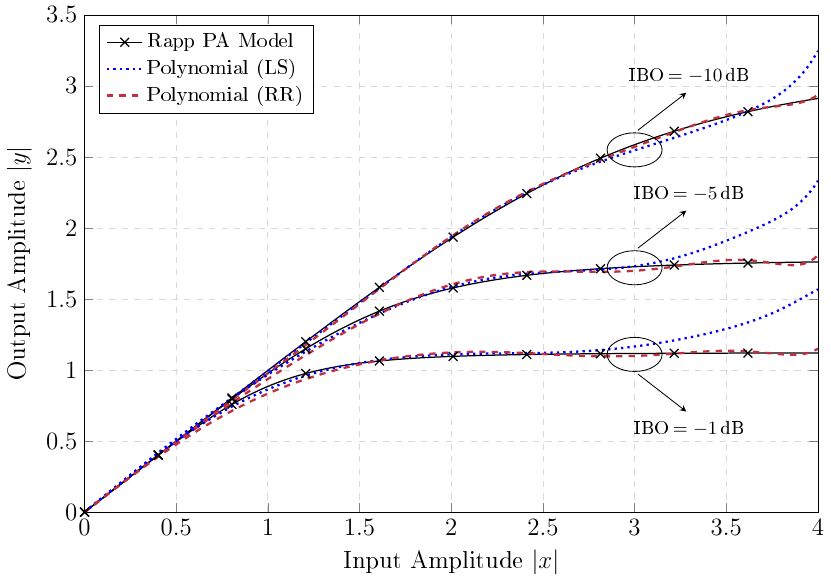}
\caption{AM-AM characteristic comparison of Rapp PA model.}
\label{fig:pa_fitting_ls}
\end{figure}

\begin{figure}[t]
    \centering
    \includegraphics[width=\columnwidth]{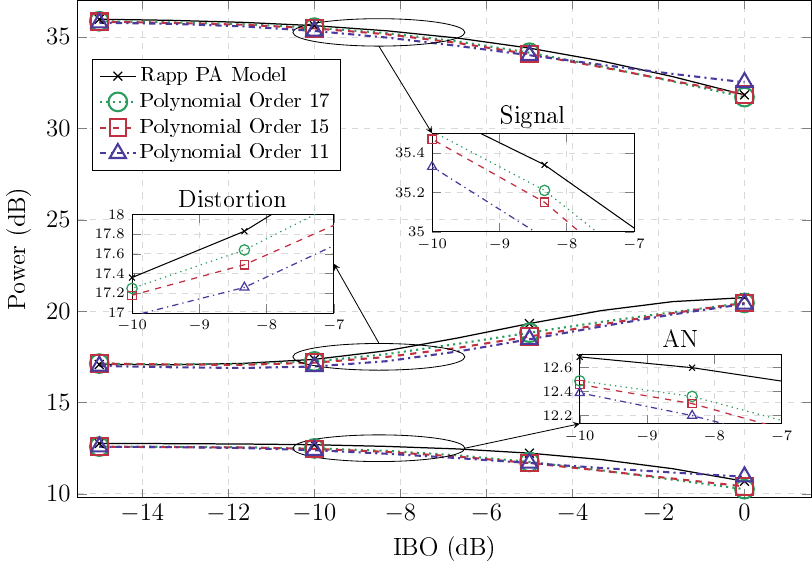}
    \caption{Validation of polynomial \gls{pa} against the Rapp model.}
    \label{fig:power_components}
\end{figure}

\begin{figure}[t]
    \centering
    \includegraphics[width=\columnwidth]{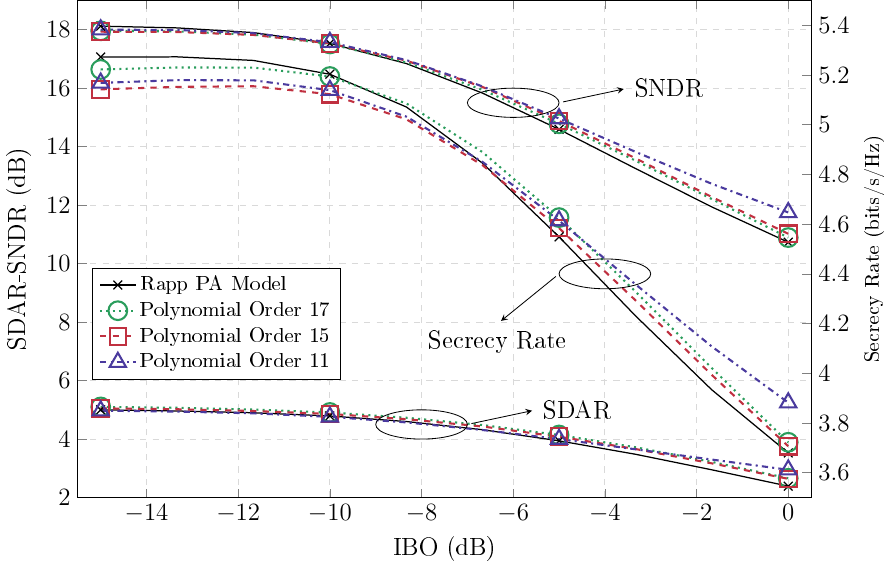}
    \caption{SDAR, SNDR, and secrecy rate (right axis) versus \gls{ibo}.}
    \label{fig:sndr_sdar_secrecy}
\end{figure}

\cref{tab:simulation_parameters} summarizes the key simulation parameters. The system model, illustrated in \cref{fig:system_model} and detailed in \cref{sec:system_model}, operates under a nonlinear \gls{pa} regime, with all secrecy rate computations following the analytical Bussgang decomposition formulas derived in \cref{sec:sndr_formulation}. \cref{fig:pa_fitting_ls} compares \gls{ls} and \gls{rr} applied to the Rapp AM-AM characteristic using 1000 uniformly spaced amplitude samples over \( |x| \in [0, \sqrt{P_t}] \); {{at order 15 under 32-bit floating-point precision, condition numbers exceed \( 10^{17} \), causing \gls{ls} to diverge from the Rapp curve while \gls{rr} remains stable, motivating the use of \gls{rr} over \gls{ls}. The polynomial order is set to 15 to balance approximation accuracy against training complexity, with detailed regularization analysis provided in~Appendix~\ref{app:appendix_float32}. We then analyze how \gls{ibo} affects key power components at the eavesdropper and legitimate user using Bussgang decomposition with \gls{mrt} precoding, and validate the polynomial approximations against the Rapp \gls{pa} model. The parameters $M = 64$, $N_e = 3$, and $P_{\text{AN}} = 30\%$ are used only for this analytical validation in \cref{fig:power_components,fig:sndr_sdar_secrecy}, while all other results use the parameters in \cref{tab:simulation_parameters}.}}
\begin{table}[t]
\centering
\caption{Simulation Parameters}
\label{tab:simulation_parameters}
\begin{tabular}{lll}
\toprule
\textbf{Symbol} & \textbf{Parameter} & \textbf{Value} \\
\midrule
\multicolumn{3}{l}{\textbf{System Configuration}} \\
\( M \) & Transmit antennas & 16 \\
\( K \) & Legitimate users & 3 \\
\( N_e \) & Eavesdroppers & 5 \\
\( P_t \) & Total transmit power & 16 W \\
\( \sigma_{\nu}^2, \sigma_{\varepsilon}^2\) & Noise variance & 0.1 W \\
\( \theta_k \) & User angles & Uniform [0, $\pi$] \\
\midrule
\multicolumn{3}{l}{\textbf{Power Amplifier}} \\
\( \text{\gls{ibo}} \) & Input back-off & $-15, -10, -5, -1$ dB \\
\( p \) & Rapp smoothness parameter & 3 \\
-- & Polynomial order & 15 \\
-- & Fitting method & Ridge regression \\
-- & Power terms & Odd only \\
\midrule
\multicolumn{3}{l}{\textbf{GNN Architecture}} \\
\( L \) & Message passing layers & 6 \\
\( d \) & Hidden dimension & 128 \\
\midrule
\multicolumn{3}{l}{\textbf{Training Configuration}} \\
\( N_{\text{train}} \) & Training samples & 350,000 \\
\( N_{\text{val}} \) & Validation samples & 25,000 \\
\( N_{\text{test}} \) & Test samples & 16,000 \\
\( N_{\text{epochs}} \) & Maximum epochs & 450 \\
\( \eta \) & Initial learning rate & $5 \times 10^{-3}$ \\
\( B \) & Batch size & 128 \\
\midrule
\multicolumn{3}{l}{\textbf{Hardware}} \\
-- & \gls{gpu} & NVIDIA Tesla T4 \\
-- & \gls{cpu} & Intel Core Ultra 7 \\
\midrule
\multicolumn{3}{l}{\textbf{Baseline Configuration}} \\
\( \alpha \) & Signal power fraction (AN) & Search over $[0.3, 1.0]$ \\
\( I \) & Optimization iterations (Opt-GNN) & 200 \\
\bottomrule
\end{tabular}
\end{table}

\cref{fig:power_components} shows the evolution of signal power, distortion power, and \gls{an} power at the worst-case eavesdropper as \gls{ibo} varies from \SI{-15}{\dB} to \SI{0}{\dB}. As \gls{ibo} increases, the \gls{pa} enters deeper saturation regions, introducing stronger nonlinear distortion. {{The linear part of the}} signal power decreases monotonically for all \gls{pa} models due to amplitude compression in saturation. Distortion power increases from approximately \SI{17}{\dB} at \gls{ibo} = \SI{-15}{\dB} to \SI{21}{\dB} at \gls{ibo} = \SI{0}{\dB}, reflecting increased saturation effects. The \gls{an} power at the eavesdropper decreases from \SI{13}{\dB} at \gls{ibo} = \SI{-15}{\dB} to \SI{10}{\dB} at \gls{ibo} = \SI{0}{\dB}, due to the interaction between \gls{an} and the Bussgang gain matrix. As shown in \cref{fig:power_components}, the polynomial approximations become closer to the Rapp model as the order increases; however, higher orders increase numerical instability and optimization complexity.

\cref{fig:sndr_sdar_secrecy} shows the SDAR at the worst-case noiseless eavesdropper and SNDR at the legitimate user as functions of \gls{ibo}. As \gls{ibo} increases toward saturation, \gls{sdar} and \gls{sndr} both decrease across all \gls{pa} models, as stronger distortion degrades eavesdropper reception, improving security, while simultaneously reducing legitimate user \gls{sndr}. These competing effects result in a monotonically decreasing secrecy rate as \gls{ibo} increases, with the highest secrecy rate achieved at linear operation and diminishing returns as saturation deepens. The figure further confirms that lower-order polynomial approximations introduce Bussgang coefficient errors that propagate into the \gls{snidr} computations and finally into the secrecy rate, while higher orders increase training complexity through more nonlinear coupling terms; a 15th-order polynomial therefore provides a suitable trade-off between approximation accuracy and computational efficiency.

\subsection{GNN Training Convergence}
\label{subsec:training_convergence}
\begin{figure}[t]
\centering
\includegraphics[width=\columnwidth]{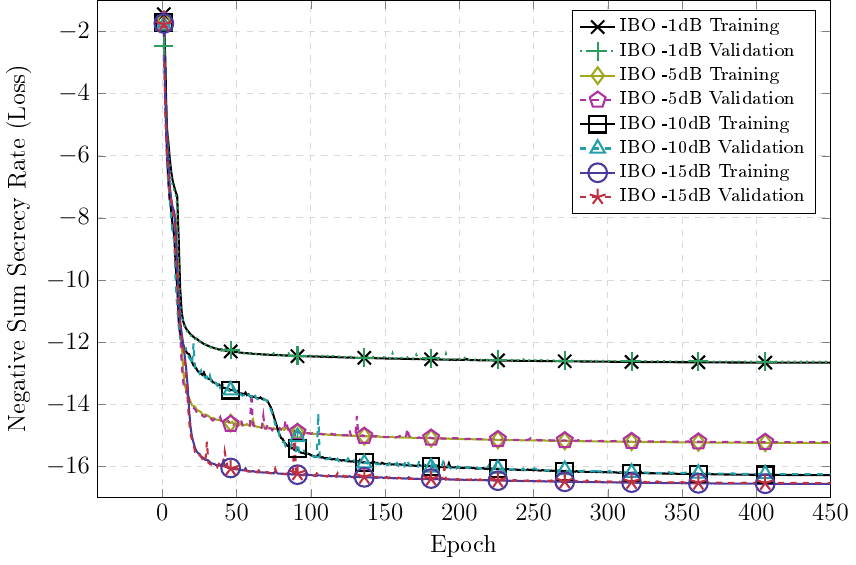}
\caption{Training and validation loss curves for various IBOs.}
\label{fig:training_curves_ibo_m1db}
\end{figure}
{{\cref{fig:training_curves_ibo_m1db} presents training and validation loss curves across four \gls{ibo} levels over 450 epochs. The validation loss closely tracks training loss, indicating good generalization. The converged sum secrecy rate decreases from \SI{16.53}{bps/Hz} at IBO = \SI{-15}{\dB}, corresponding to linear operation, to \SI{16.27}{bps/Hz}, \SI{15.23}{bps/Hz}, and \SI{12.66}{bps/Hz} at IBO = \SI{-10}{\dB}, \SI{-5}{\dB}, and \SI{-1}{\dB}, respectively. This 23\% performance reduction under severe saturation at IBO = \SI{-1}{\dB} compared to linear operation has two contributing factors. First, stronger \gls{pa} distortion fundamentally lowers the achievable secrecy rate, since distortion increases the interference in the legitimate user \gls{snidr}, and spatially shaping distortion away from users toward eavesdroppers consumes array gain, and the deeper the saturation the more antennas and power are needed to shape the distortion, so the optimal secrecy rate is inherently lower regardless of precoder quality. Second, stronger nonlinearity introduces a more non-convex optimization landscape, making it additionally harder for the \gls{gnn} to approach even that lower optimum.}}

\label{subsec:performance_comparison}
\begin{figure*}[!t]
\centering
\includegraphics[width=0.85\textwidth]{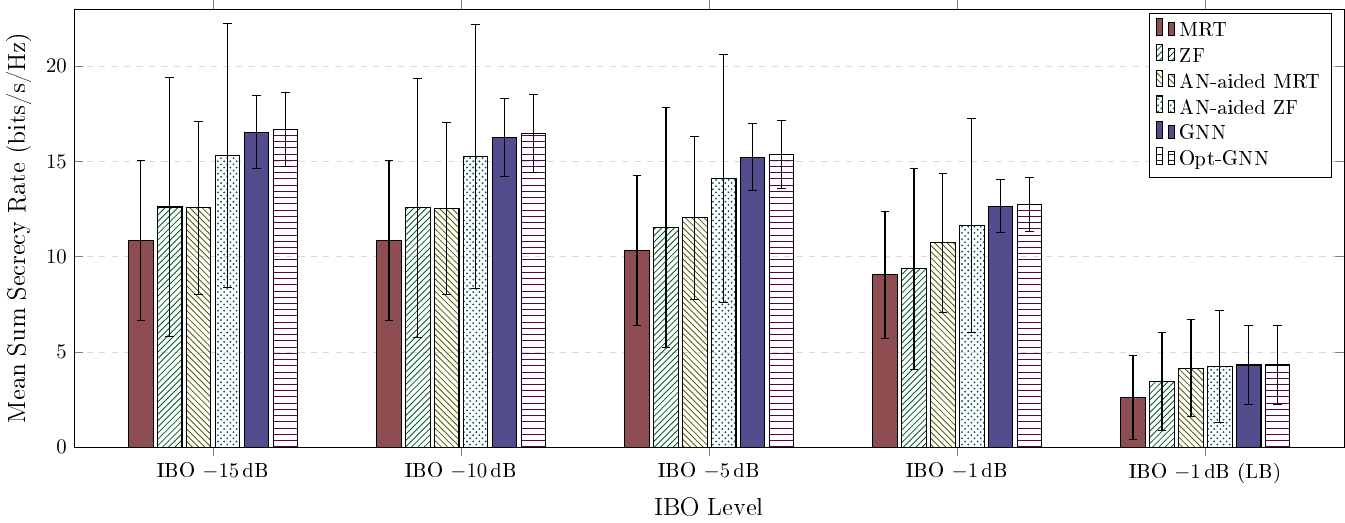}
\caption{Performance comparison across \gls{ibo} levels with error bars.}
\label{fig:performance_comparison_all_ibo}
\end{figure*}

\cref{fig:performance_comparison_all_ibo} presents a performance comparison across four \gls{ibo} levels ranging from linear operation at \SI{-15}{\dB} to severe saturation at \SI{-1}{\dB}. {{Throughout this section, the reported mean and standard deviation are computed over independent channel realizations.}} All methods exhibit decreasing sum secrecy rates as \gls{ibo} increases toward saturation. As discussed in \cref{subsec:training_convergence}, this reflects both a fundamentally lower achievable secrecy rate due to stronger \gls{pa} distortion and a more non-convex optimization landscape under deeper saturation.

At severe saturation with IBO = \SI{-1}{\dB}, where distortion is strongest and the optimization landscape most non-convex, the \gls{gnn} achieves \SI{12.66}{bps/Hz}, outperforming \gls{mrt} at \SI{9.05}{bps/Hz}, \gls{zf} at \SI{9.36}{bps/Hz}, AN-aided MRT at \SI{10.73}{bps/Hz}, and AN-aided ZF at \SI{11.65}{bps/Hz}. The gain over all baselines reflects the \gls{gnn}'s ability to exploit the spatial structure of \gls{pa} distortion, which distortion-blind precoders cannot leverage. The Opt-GNN baseline reaches \SI{12.74}{bps/Hz}, a marginal improvement over the \gls{gnn} that confirms the \gls{gnn} output is already near the local optimum of the gradient ascent refinement. This advantage persists at all operating points. At moderate saturation with IBO = \SI{-10}{\dB}, the \gls{gnn} achieves \SI{16.27}{bps/Hz}, and at linear operation with IBO = \SI{-15}{\dB}, it reaches \SI{16.53}{bps/Hz}, while the strongest baseline, AN-aided ZF, reaches only \SI{15.33}{bps/Hz}.

The \gls{gnn} maintains consistently low variance across all conditions with standard deviation below \SI{2.05}{bps/Hz}, whereas \gls{zf}-based methods exhibit standard deviation of at least \SI{5.30}{bps/Hz} due to sensitivity to channel conditioning, corresponding to a 58.13--75.31\% reduction in standard deviation over all baselines at IBO~$= -1$~dB. This stability ensures reliable, secure communications across diverse channel realizations. Taken together with its superior mean secrecy rate, the \gls{gnn} is the only method that simultaneously occupies the high-mean, low-variance region across all \gls{ibo} levels, a joint advantage that no baseline achieves at any operating point.

These results demonstrate that \gls{pa} distortion can be exploited to enhance \gls{pls} when properly directed, as the \gls{gnn} outperforms \gls{mrt} by 39.89\%, \gls{zf} by 35.26\%, AN-aided MRT by 17.99\%, and AN-aided ZF by 8.67\% at severe saturation, where distortion is strongest. \cref{tab:performance_comparison} reports the \gls{gnn} gain in mean and standard deviation reduction relative to each baseline across all \gls{ibo} levels, including the worst-case lower bound at \gls{ibo}~$=-1$~dB.

\begin{figure}[t]
\centering
\includegraphics[width=\columnwidth]{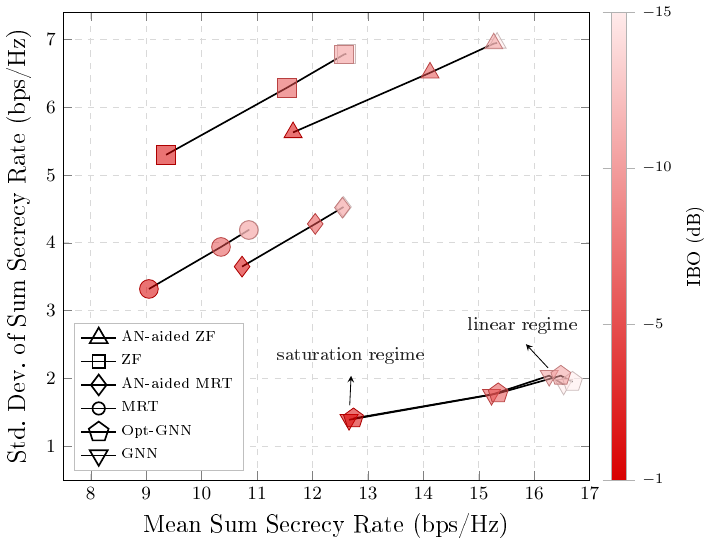}
\caption{Mean versus standard deviation of sum secrecy rate.}
\label{fig:efficiency_frontier}
\end{figure}

\cref{fig:efficiency_frontier} plots the mean sum secrecy rate against its standard deviation for each method and \gls{ibo} level, providing a joint reliability--performance view. As saturation increases, the \gls{gnn} redirects \gls{pa} distortion toward eavesdroppers rather than treating it as an impairment, which allows it to maintain both a high mean secrecy rate and a low variance across all \gls{ibo} levels. The connecting lines show that the \gls{gnn} path is short and nearly horizontal as \gls{ibo} moves toward deeper saturation, confirming that exploiting distortion produces consistent gains at every operating point. In contrast, \gls{zf}-based methods follow longer and steeper paths, reflecting their sensitivity to increasing \gls{pa} nonlinearity.
\begin{figure*}[t]
\centering
\includegraphics[width=1\textwidth]{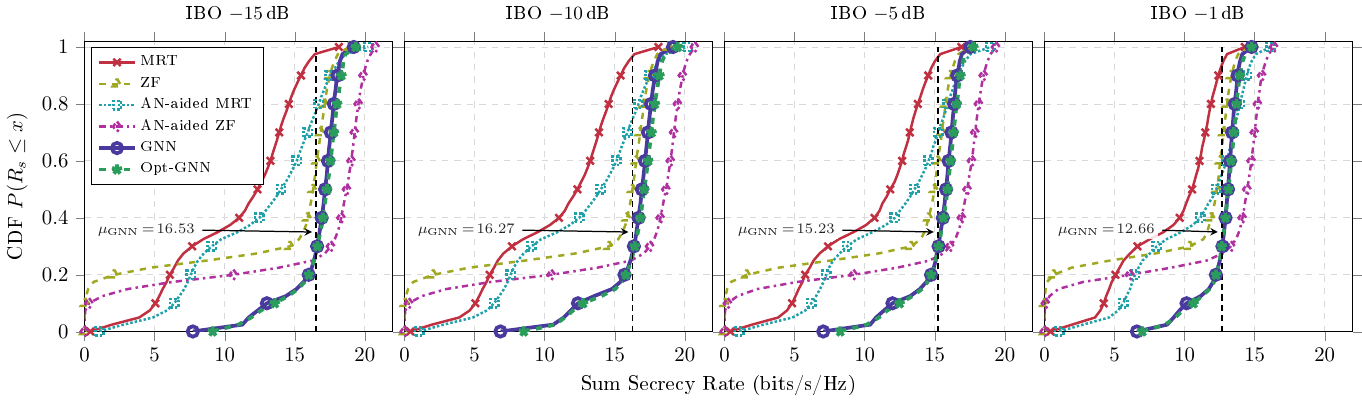}
\caption{Empirical CDF of the sum secrecy rate for all methods across four \gls{ibo} levels.}
\label{fig:cdf_combined_all_ibo}
\end{figure*}
\cref{fig:cdf_combined_all_ibo} shows the empirical \gls{cdf} across all four \gls{ibo} levels, where the \gls{gnn} achieves the narrowest distribution with a secrecy rate floor above \SI{6}{bps/Hz} even at \gls{ibo} = \SI{-1}{\dB}, while \gls{zf}-based methods exhibit lower tails with near-zero secrecy in a significant fraction of realizations. {{In the upper tail, AN-aided \gls{zf} occasionally surpasses the \gls{gnn} for favorable channels, but the \gls{gnn}'s much higher floor and narrower spread remain more important for reliable secrecy.}}

\subsection{Robustness to Eavesdropper Count}
\label{subsec:robustness_eval}
{
{\cref{fig:robustness_eavesdroppers} evaluates generalization across varying eavesdropper counts. Each data point reports the mean sum secrecy rate over 1000 independent test samples, where each sample draws a fresh eavesdropper realization according to  \cref{subsec:channel_models}, i.e., a new random angular shift \( \delta_s \sim \mathcal{U}(0, \pi/(N_e-1)) \) per sample; shaded bands show the standard deviation normalized by \( \sqrt{1000} \). At IBO = \SI{-1}{\dB}, the \gls{gnn} maintains a consistent mean sum secrecy rate from \SIrange{12.66}{12.88}{bps/Hz} across \( N_e \in \{3, 5, 7, \ldots, 15\} \). This consistency arises because the loss function averages the sum secrecy rate over eavesdropper realizations, so the mean remains stable as \( N_e \) grows.
\begin{figure}[t]
\centering
\includegraphics[width=\columnwidth]{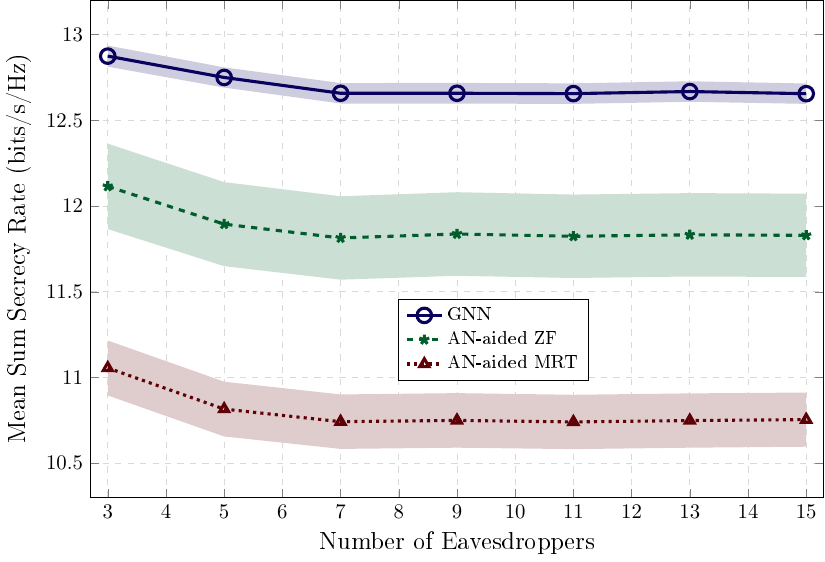}
\caption{Mean sum secrecy rate versus \( N_e \) at IBO = \SI{-1}{\dB}.}
\label{fig:robustness_eavesdroppers}
\end{figure}
The \gls{gnn} consistently outperforms AN-aided MRT and AN-aided ZF by 8.67\% at high saturation, with the advantage decreasing to 7--8\% at moderate saturation, confirming that the model has learned generalizable precoding strategies rather than eavesdropper-specific patterns.
}}
\begin{table}[t]
\centering
\caption{GNN gain $(\Delta\mu\%,\,\Delta\sigma\%)$ over baselines.}
\label{tab:performance_comparison}
\renewcommand{\arraystretch}{1.15}
\setlength{\tabcolsep}{4pt}
\footnotesize
\begin{tabular}{c|c|c|c}
\toprule
\shortstack{IBO\\(dB)} & vs MRT & vs ZF & vs AN-ZF \\
\midrule
$-15$ & $(+52.21,\,+54.42)$ & $(+31.09,\,+71.87)$ & $(+7.83,\,+72.52)$ \\
$-10$ & $(+49.95,\,+51.31)$ & $(+29.44,\,+69.91)$ & $(+6.55,\,+70.61)$ \\
$-5$  & $(+47.15,\,+55.33)$ & $(+31.98,\,+72.02)$ & $(+7.86,\,+72.96)$ \\
$-1$  & $(+39.89,\,+58.13)$ & $(+35.26,\,+73.77)$ & $(+8.67,\,+75.31)$ \\
\midrule
$-1^{\,\dagger}$ & $(+65.13,\,+4.13)$ & $(+24.57,\,+18.36)$ & $(+1.65,\,+28.91)$ \\
\bottomrule
\multicolumn{4}{l}{\scriptsize $^\dagger$ Worst-case lower bound~(\cref{subsec:secrecy_rate_lower_bound}).}
\end{tabular}
\end{table}

%% file: sections/07_complexity.tex
\section{Computational Complexity and Inference Time}
\label{sec:complexity}
{{Complexity is measured in \glspl{flop}, where one \gls{flop} is one real multiply or add~\cite{feys2025energy}. \cref{tab:complexity} summarizes the complexity, example \glspl{flop}, and \gls{cpu} inference time per channel realization for all methods at \( \mathrm{IBO}\!=\!{-1}\,\mathrm{dB} \). The \gls{gnn} forward pass executes \( \mathcal{O}(LMKd(d+M+K)) \) \glspl{flop} across the \( M\cdot K \) edges, from edge updates and aggregation at antenna and user nodes, around \( 5\times 10^{6} \) \glspl{flop} for our setup. Since edge updates are independent, they can be parallelized across the \( M\cdot K \) edges, reducing the work per edge to \( \mathcal{O}(L(d^2 + Md + Kd)) \), about \( 1\times 10^{5} \) \glspl{flop} on parallel hardware. The \gls{cpu} time of \SI{0.70}{\milli\second} in \cref{tab:complexity} corresponds to the full \( 5\times 10^{6} \) \glspl{flop}, since the \gls{cpu} executes the forward pass sequentially. Training is performed offline on a \gls{gpu} and is not part of the real-time inference budget.}}

\begin{table}[t]
\caption{Method complexity and inference time.}
\label{tab:complexity}
\centering
\begin{tabular}{lccc}
\toprule
\textbf{Method} & \textbf{Complexity} & \textbf{\(\sim\)FLOPs} & \textbf{Time (ms)} \\
\midrule
MRT             & \( \mathcal{O}(MK) \)                & \(\sim\)5e1 & 0.006 \\
ZF              & \( \mathcal{O}(MK^2\!+\!K^3) \)     & \(\sim\)2e2 & 0.016 \\
AN-aided MRT    & \( \mathcal{O}(M^3) \)               & \(\sim\)4e3 & 0.055 \\
AN-aided ZF     & \( \mathcal{O}(M^3) \)               & \(\sim\)4e3 & 0.066 \\
GNN             & \( \mathcal{O}(LMKd(d\!+\!M\!+\!K)) \) & \(\sim\)5e6 & 0.70 \\
Opt-GNN         & \( \mathcal{O}(IM^2KN_e) \)          & \(\sim\)8e5 & 1923 \\
\bottomrule
\end{tabular}
\end{table}

The iterative optimization applies \( I\!=\!200 \) gradient-ascent steps per channel realization, where each step evaluates the \gls{snidr} for \( K \) users and \( KN_e \) eavesdroppers; computing \( \mat{C}_e(\mat{C}_x) \) and quadratic forms \( \vect{h}^H\mat{C}_e\vect{h} \) costs \( \mathcal{O}(M^2) \) per channel vector, giving \( \mathcal{O}(IM^2KN_e) \) total.
Although Opt-GNN has fewer \glspl{flop} than the \gls{gnn}, its iterations must run one after another, giving a measured inference time of \SI{1923}{\milli\second} versus \SI{0.70}{\milli\second} for the \gls{gnn} on an Intel Core Ultra 7 155H \gls{cpu}, a \(\sim\!2740\times\) speedup. This shows that \gls{flop} count alone does not reflect real inference speed when parallelism differs across methods.
Compared to classical and \gls{an}-aided methods, the \gls{gnn} takes  longer (\SI{0.70}{\milli\second} vs.\ \(\leq\)\SI{0.066}{\milli\second}) but achieves higher secrecy rates without per-sample optimization or \gls{an} power allocation.
The \SI{0.70}{\milli\second} accounts only for precoding computation; the \gls{3gpp} \gls{embb} user plane latency budget of \SI{4}{\milli\second}~\cite{3gpp_tr38913} covers the full transmission pipeline including scheduling, channel estimation, and propagation, so precoding represents a small fraction of that budget.

%% file: sections/08_conclusion.tex
\section{Conclusion}
\label{sec:conclusion}
This paper presented a \gls{gnn}-based precoding framework for \gls{pls} in multi-user \gls{miso} systems with nonlinear power amplifiers. The proposed approach exploits \gls{pa} distortion as a beneficial mechanism rather than treating it as an impairment to be compensated. The bipartite \gls{gnn} architecture captures antenna-user relationships through iterative message passing and learns to maximize sum secrecy rates directly from legitimate user channel state information. High-order polynomial with ridge regression ensures accurate \gls{pa} characterization at low \gls{ibo} values, while Bussgang decomposition with factorial-based formulas enables efficient analytical gradient computation for training. Numerical evaluation demonstrates that the \gls{gnn} achieves \SI{12.66}{bps/Hz} at severe saturation with IBO = \SI{-1}{\dB}, outperforming \gls{mrt} by 39.89\%, \gls{zf} by 35.26\%, AN-aided MRT by 17.99\%, and AN-aided ZF by 8.67\%, while performing within 0.63\% of iterative optimization. {{The absolute secrecy rate improves as the \gls{pa} moves away from saturation; the \gls{gnn}'s relative advantage over baselines narrows in the linear regime, where classical methods also perform well, but the \gls{gnn} still achieves the highest mean and lowest variance across all \gls{ibo} levels.}} The approach maintains robust performance across \gls{ibo} levels from \SI{-15}{\dB} to \SI{-1}{\dB} and generalizes to varying eavesdropper counts with consistent secrecy rate of \SI{12.66}{bps/Hz} across \( N_e \in \{3, \ldots, 15\} \). 
Future work includes extension to general fading channels, imperfect  \gls{csi} scenarios, multi-cell deployments with inter-cell interference, and measurement-based validation on hardware testbeds.

%% file: sections/09_appendix.tex
\section{Loss Function Analysis with MRT}
\label{app:appendix_loss_derivation}
This section analyzes a specific case with a single legitimate user and linear \gls{pa} regime to {{provide intuition that}} \gls{mrt} does not maximize the expected secrecy rate \( \mathbb{E}_{\vect{h}_E}[f(\vect{w})] \), where \( f(\vect{w}) = \log_2\!\bigl((1 + \text{SNR}_L)/(1 + \text{SNR}_E)\bigr) \).
{{The approximation \( \mathbb{E}[\log(x/y)] \approx \log(\mathbb{E}[x]/\mathbb{E}[y]) \) becomes more accurate as the variance of $x/y$ decreases relative to its mean; the derived bounds should therefore be interpreted as approximate rather than exact.}} The gradient of \( f(\vect{w}) \) becomes
\( \nabla_{\vect{w}} f(\vect{w}) \approx \nabla A(\vect{w})/A(\vect{w}) - \nabla B(\vect{w})/B(\vect{w}) \),
where \( A(\vect{w}) = \sigma^2 + \vect{w}^H \vect{h}_L \vect{h}_L^H \vect{w} \) and
\( B(\vect{w}) = \sigma^2 + \vect{w}^H \mathbb{E}[\vect{h}_E \vect{h}_E^H] \vect{w} \). For a two-antenna system with \( \vect{h}_L = [1, e^{-j\phi}]^T \), \( \vect{h}_E(\theta) = [1, e^{-j\theta}]^T \), and \( \theta \sim \text{Uniform}[0, \pi] \), the \gls{mrt} precoder is \( \vect{w}_{\text{MRT}} = [1, e^{j\phi}]^T/\sqrt{2} \).
The expected eavesdropper covariance is \( \mathbb{E}[\vect{h}_E \vect{h}_E^H] = \bigl[\begin{smallmatrix} 1 & 2j/\pi \\ -2j/\pi & 1 \end{smallmatrix}\bigr] \), since \( \mathbb{E}[e^{j\theta}] = \frac{1}{\pi}\int_0^{\pi} e^{j\theta}\, d\theta = 2j/\pi \). Evaluating at \gls{mrt}, \( A(\vect{w}_{\text{MRT}}) = \sigma^2 + 1 + \cos(2\phi) \),
\( B(\vect{w}_{\text{MRT}}) = \sigma^2 + 1 - \frac{2\sin\phi}{\pi} \),
\( \nabla_{\vect{w}} A|_{\text{MRT}} = 2\sqrt{2}\cos\phi\,e^{j\phi} [1, e^{-j\phi}]^T \), and
\( \nabla_{\vect{w}} B|_{\text{MRT}} = \sqrt{2} [1 + \frac{2j}{\pi} e^{j\phi}, -\frac{2j}{\pi} + e^{j\phi}]^T \).
The full gradient at \gls{mrt} is
\begin{equation}
\begin{split}
\nabla_{\vect{w}} f(\vect{w})\big|_{\text{MRT}} \approx \sqrt{2} \bigg[ & \frac{2\cos\phi\,e^{j\phi}[1, e^{-j\phi}]^T}{\sigma^2 + 1 + \cos(2\phi)} \\
& - \frac{[1 + \frac{2j}{\pi} e^{j\phi}, -\frac{2j}{\pi} + e^{j\phi}]^T}{\sigma^2 + 1 - \frac{2\sin\phi}{\pi}} \bigg]
\end{split}
\end{equation}
For the gradient to be zero, the following condition must hold
\begin{equation}
\frac{2[1, e^{-j\phi}]^T}{\sigma^2 + 1 + \cos(2\phi)} = \frac{[1 + \frac{2j}{\pi}e^{j\phi}, -\frac{2j}{\pi} + e^{j\phi}]^T}{\sigma^2 + 1 - \frac{2\sin\phi}{\pi}}
\end{equation}
{{This condition cannot be satisfied for most \( \phi \) and \( \sigma^2 \), suggesting
that \gls{mrt} is not a stationary point of the expected secrecy rate under this approximation. This motivates the use of
\gls{gnn} to learn precoders that account for the statistical distribution of the eavesdropper
channels.}}

\section{Ridge Regression Accuracy}
\label{app:appendix_float32}
For 32-bit floating-point precision, the design-matrix normal-equations condition number \( \kappa(\mat{X}^H \mat{X}) \) grows rapidly with polynomial order, from \( 3.1\times 10^{15} \) at order 13, to \( 2.2\times 10^{18} \) at order 15, and \( 1.8\times 10^{20} \) at order 17. These values exceed the 32-bit numerical stability threshold (\( \sim 10^{7} \)), requiring regularization. The optimal \( \alpha^{\star} \) varies by IBO level. \cref{tab:float32_regularization} reports the optimal values obtained via grid search over \( \alpha \in [0, 10] \) at order 15 under 32-bit floating-point precision; the IBO grid matches \cref{tab:simulation_parameters}. The MSE is computed as \( \frac{1}{N}\sum_{k=1}^{N}|y_{\text{Rapp}}(A_k) - y_{\text{poly}}(A_k)|^2 \) over \( N \) test amplitudes \( A_k \) uniformly sampled in \( [0, \sqrt{P_t}] \).
\begin{table}[ht]
\centering
\footnotesize
\caption{\gls{rr} regularization parameter \( \alpha^{\star} \) at order 15.}
\label{tab:float32_regularization}
\begin{tabular}{cccc}
\toprule
IBO (dB) & $\alpha^{\star}$ & MSE (64-bit) & MSE (32-bit) \\
\midrule
-15 & 4.46        & 1.92e-4 & 1.78e-3 \\
-10 & 1.61        & 1.85e-4 & 1.83e-3 \\
-5  & 3.81        & 2.10e-4 & 3.27e-3 \\
-1  & 6.45e-2     & 2.39e-4 & 3.42e-3 \\
\bottomrule
\end{tabular}
\end{table}
With tuned \( \alpha \), 32-bit precision achieves MSE \( \leq 4 \times 10^{-3} \), comparable to the 64-bit baseline.